%% file: aaai22.tex
\def\year{2022}\relax
\documentclass[letterpaper]{article} 
\usepackage{aaai22}  
\usepackage{times}  
\usepackage{helvet}  
\usepackage{courier}  
\usepackage[hyphens]{url}  
\usepackage{graphicx} 
\usepackage{natbib}  
\usepackage{caption} 
\DeclareCaptionStyle{ruled}{labelfont=normalfont,labelsep=colon,strut=off} 
\usepackage{algorithm}
\usepackage{algorithmic}

\usepackage{newfloat}
\usepackage{listings}
\floatstyle{ruled}
\newfloat{listing}{tb}{lst}{}
\floatname{listing}{Listing}
\title{Leveraging Experience in Lifelong Multi-Agent Pathfinding}
\author{
    Nitzan Madar,
    Kiril Solovey,
    Oren Salzman
}
\affiliations{

    Technion, Israel Institute of Technology\\
    nitzanmadar@campus.technion.ac.il,
    kirilsol@technion.ac.il, 
    osalzman@cs.technion.ac.il
}

\usepackage{bibentry}

\input{tex/macros}

\usepackage{hyperref}
\usepackage{color}
\usepackage{xspace}
\usepackage{amsmath}
\usepackage{import,svg} 
\usepackage{subcaption} 
\usepackage{amsfonts} 
\usepackage{amsthm}
\usepackage[binary-units=true]{siunitx}

\begin{document}

\maketitle

\begin{abstract}

In Lifelong Multi-Agent Path Finding (L-MAPF) 
a team of agents performs a stream of tasks consisting of multiple locations to be visited by the agents on a shared graph while avoiding collisions with one another. L-MAPF is typically tackled by partitioning it into \emph{multiple} \emph{consecutive}, and hence \emph{similar}, ``one-shot'' MAPF queries, 
as in the Rolling-Horizon Collision Resolution (\rhcr) algorithm. Therefore, a solution to one query informs the next query, which leads to similarity with respect to the agents' start and goal positions, and how collisions 
need to be resolved from one query to the next. Thus, \emph{experience} from solving one MAPF query can potentially be used to speedup solving the next one. 
Despite this intuition, current L-MAPF planners solve consecutive MAPF queries \emph{from scratch}. 
In this paper, we introduce a new \rhcr-inspired approach called \exrhcr, which exploits experience in its constituent MAPF queries. In particular, \exrhcr employs an extension of Priority-Based Search (\pbs), a state-of-the-art MAPF solver. The extension, which we call \expbs, allows to \emph{warm-start} the search with the priorities between agents used by  \pbs in the previous MAPF instances.
We demonstrate empirically that \exrhcr solves L-MAPF instances up to 39\% faster than \rhcr, and has the potential to increase system  throughput for given task streams by increasing the number of agents a planner can cope with for a given time budget.
\end{abstract}

\graphicspath{{graphics/}}

\input{tex/intro_and_background}

\input{tex/probelm_definition}

\input{tex/algorithmic_background}

\input{tex/algorithm}

\input{tex/empirical_results}

\input{tex/conclusions}

\input{tex/acknowledgements}

\bibliography{aaai22}

\ifthenelse{\boolean{arXiv}}{
    \input{tex/appendix-new}

}
{
}

\end{document}

%% file: tex/macros.tex
\newcommand{\calA}{\ensuremath{\mathcal{A}}\xspace}

\newcommand{\calP}{\ensuremath{\mathcal{P}}\xspace}
\newcommand{\calT}{\ensuremath{\mathcal{T}}\xspace}

\newcommand\algname[1]{\textsf{#1}\xspace}
\newcommand\astar{\algname{A*}}

\newcommand\cbs{\algname{CBS}}
\newcommand\pbs{\algname{PBS}}
\newcommand\dfs{\algname{DFS}}

\newcommand\wldfs{\algname{WL\nobreakdash-DFS}}

\newcommand\expbs{\algname{exPBS}}

\newcommand\rhcr{\algname{RHCR}}
\newcommand\exrhcr{\algname{exRHCR}}
\newcommand{\ignore}[1]{}
\newcommand\quot[1]{``{#1}''\xspace}

\newcommand{\PREC} {\boldsymbol{\pmb\prec}}
\newcommand{\PRECexp} {$\PREC_{\text{exp}}$\xspace}

\newcommand{\Cpp}{C\raise.08ex\hbox{\tt ++}\xspace}

\newcommand{\warehouse}{\textsc{Warehouse}\xspace}
\newcommand{\sorting}{\textsc{Sorting}\xspace}

\newcommand{\wmapf}{W\nobreakdash-MAPF\xspace}

\newboolean{HIDENOTES}
\setboolean{HIDENOTES}{false}  

\ifthenelse{\boolean{HIDENOTES}}{  
    \newcommand{\OS}[1]{{}}
    \newcommand{\NM}[1]{{}}
    \newcommand{\KS}[1]{{}}
    \newcommand{\bookmark}[1]{{}}
    } 
    { 
     
    \newcommand{\OS}[1]{\textcolor{blue}{#1}}
    \newcommand{\NM}[1]{\textcolor{red}{#1}}
    \newcommand{\KS}[1]{\textcolor{purple}{#1}}
    
    \newcommand{\bookmark}[1]{\colorbox{orange}{\textcolor{blue}{\LARGE{$\rightarrow$I'm Here! $\leftarrow$#1 \linebreak}}}}
    } 

\newboolean{arXiv}
\setboolean{arXiv}{true}  
\ifthenelse{\boolean{arXiv}}{  
    \newcommand{\myappendix}{{appendix}\xspace}
    \newcommand{\Sec}{{Section~}}
    \newcommand{\Fig}{{Figure~}}
    \newcommand{\Alg}{{Algorithm~}}

    } 
    { 
    \newcommand{\myappendix}{{extended version~\cite{Madar.ea.22}}\xspace}
    \newcommand{\Sec}{{Section~}}
    \newcommand{\Fig}{{Figure~}}
    \newcommand{\Alg}{{Algorithm~}}

    } 

%% file: tex/intro_and_background.tex
\section{Introduction and Related Work}


\begin{figure*}[t]  
\centering
    \hspace*{\fill}
    \begin{subfigure}[t]{0.58\linewidth}
        \centering
        \includegraphics[width=0.925\linewidth]{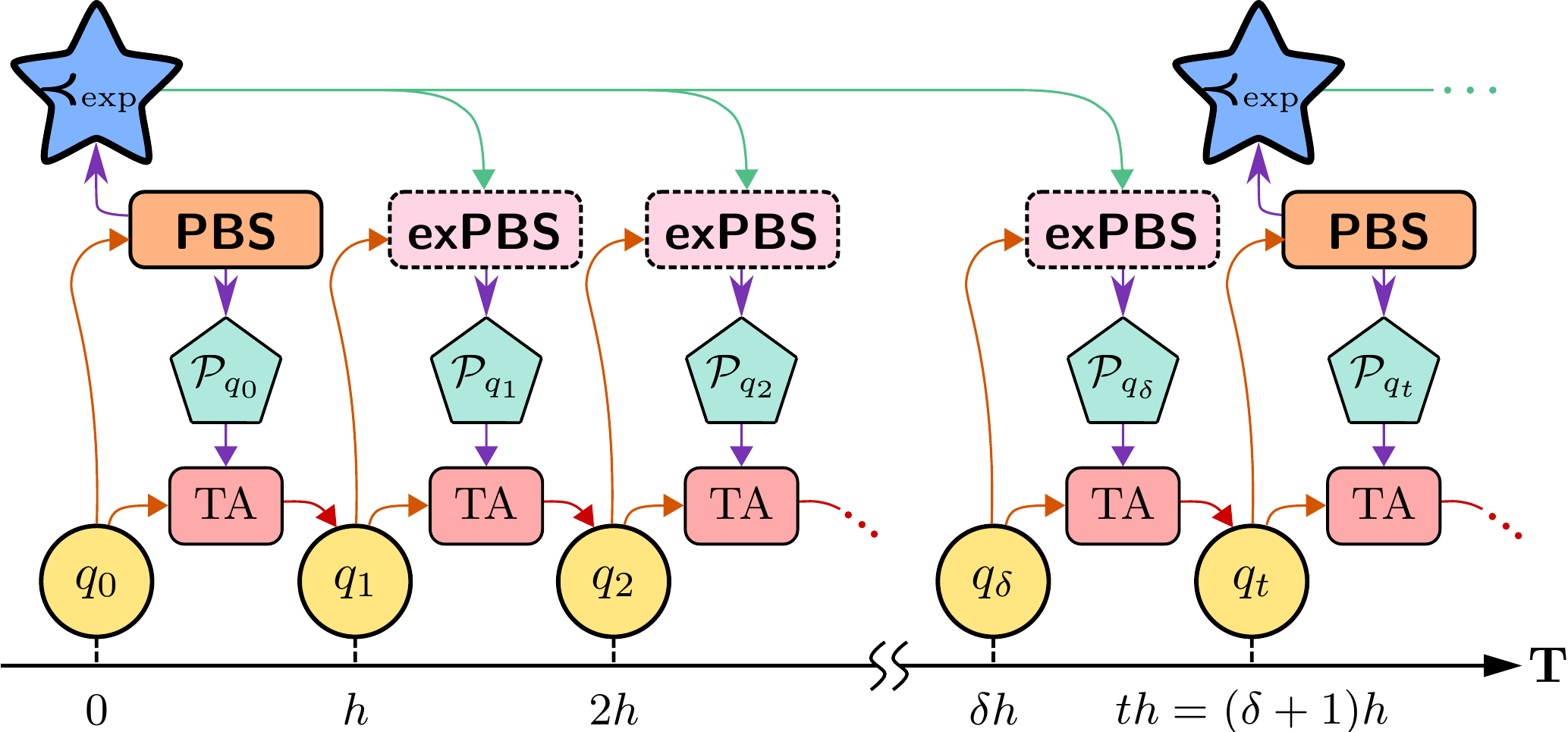}
        \caption{The \exrhcr algorithm for L-MAPF.
        } 
        \label{fig:timeline}
    \end{subfigure}
    \hspace*{\fill}
        \begin{subfigure}[t]{0.27\linewidth}
        \centering
        \includegraphics[width=0.925\linewidth]{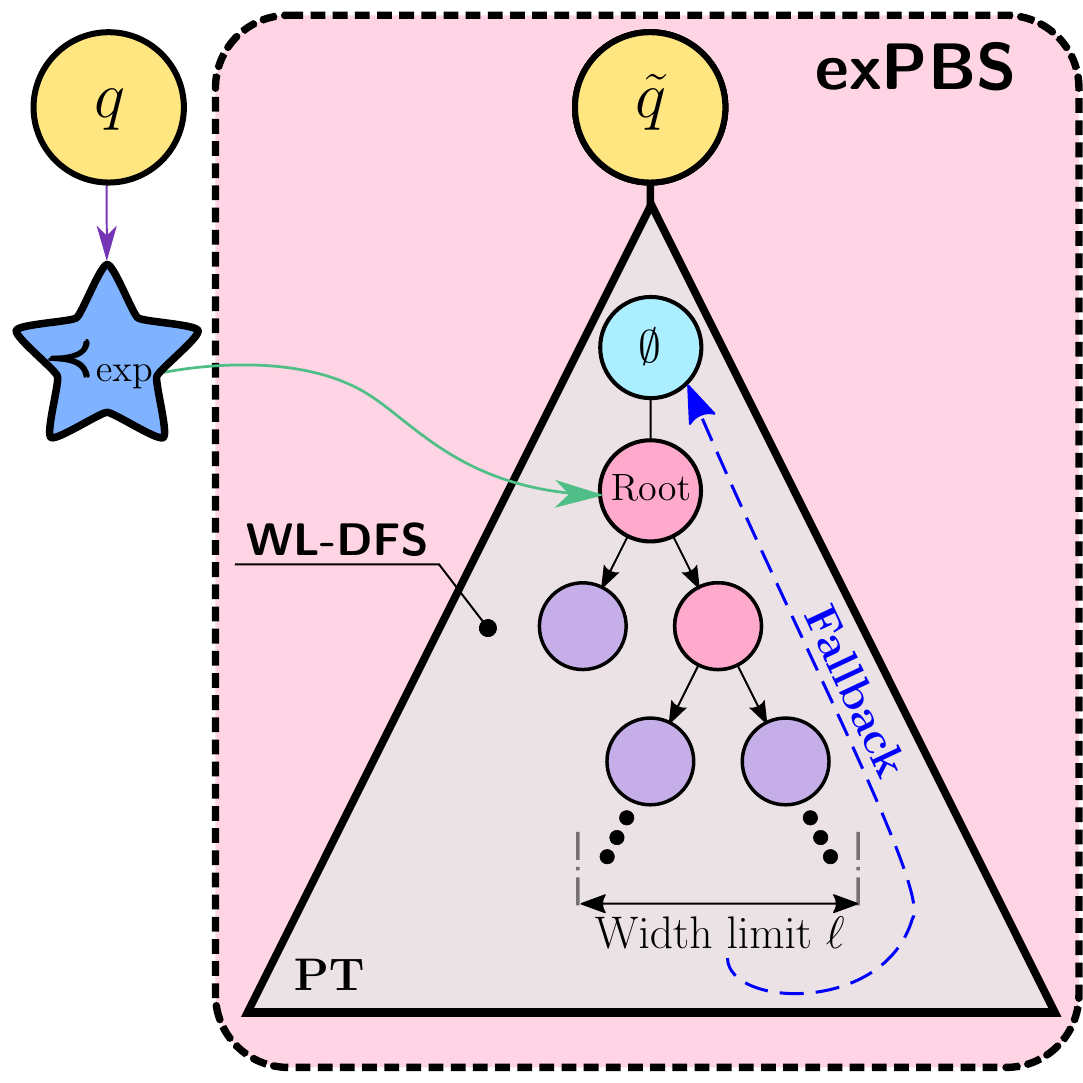}
        \caption{The \expbs algorithm for MAPF.} 
        \label{fig:expbs_changes}
    \end{subfigure}
    \hspace*{\fill}
\caption{
     Visualization of our algorithmic framework.
    \protect (\subref{fig:timeline}) To solve the L-MAPF problem, \exrhcr solves a sequence of (bounded horizon) MAPF queries $q_0, q_1, \ldots$ where a new query is solved every $h$ timesteps.
    The task assigner (TA) generates the next query~$q_{i+1}$ from a given query $q_i$ and the solution $\calP_{q_i}$ computed for $q_i$. 
    In \exrhcr 
    queries are grouped into batches of $\delta+1$ queries, where the first query in each batch is solved using \pbs whereas the next~$\delta$ queries are solved using \expbs. This is in contrast to \rhcr, which uses only \pbs. 
    The \pbs priority set \PRECexp that was used to solve $q_0$ is extracted  and used to initialize \expbs in the next~$\delta$ queries. 
    This process repeats iteratively.
    %
    %
    \protect (\subref{fig:expbs_changes})
    The internal MAPF solver \expbs leverages experience \PRECexp by using it in the root node of the priority tree (PT) to constrain and guide the search. \expbs 
    uses \textsf{WL\protect\nobreakdash-DFS}, which is a \dfs variant that limits the search tree width to avoid 
    searching in over-constrained trees. When the width limit is violated or the root PT node contains an infeasible seed priority, the original \pbs 
    is used as a fallback.
    }
    \label{fig:page1figure}
\end{figure*}

\emph{Multi-Agent Path Finding} (MAPF) is the problem of finding collision-free paths for a fleet of agents operating on a shared graph~\cite{stern2019benchmarks}. MAPF has been widely used in modeling a variety of applications including autonomous warehouse management~\cite{honig2019persistent}, 
multi-robot motion planning~\cite{Dayan.ea.21}, and multi-drone delivery~\cite{Choudhury.ea.21}.

A significant body of work is devoted to the study of the one-shot version of MAPF wherein each agent needs to reach a specific goal location from a given start. 
It is usually desirable to find \emph{high-quality solutions}, which minimize a given objective function such as the total travel time of all the agents. Finding optimal solutions to MAPF is typically computationally hard~\cite{Yu.16,Ha.ea.16}.
Nevertheless, numerous approaches have been developed which strive to return optimal or bounded-suboptimal solutions, including 
SAT solvers~\cite{Surynek.ea.16}, graph-theoretic approximation algorithms~\cite{Yu.16,Demaine.ea.19}, and search-based approaches~\cite{sharon2013increasing,wagner2015subdimensional,li2019disjoint}.

One of the most popular 
approaches for MAPF is Conflict-Based Search \textsf{(\cbs)}~\cite{sharon2015conflict}, which is a two-level optimal solver. The top level performs a best-first search on a binary \emph{constraint tree} (CT), where a given tree node specifies temporal and spatial constraints between agents in order to avoid collisions. The bottom level searches for single-agent paths that abide to the constraints in the given CT node. 
The high-level search continues until a feasible (collision-free) plan is found. 
Due to the possibly long runtime of the approach~\cite{Gordon.ea.21}, subsequent work introduced various suboptimal \cbs extensions to improve runtime~\cite{barer2014suboptimal,li2021eecbs}. A recent approach called Priority-Based Search \textsf{(\pbs)} trades off the completeness guarantees of \cbs with improved efficiency, by exploring a binary \emph{priority tree} (PT), whose nodes specify priorities between agents~\cite{ma2019searching}. Priorities can be viewed as a coarser and more computationally-efficient alternative to \cbs's constraints (see more details in \Sec \ref{sec:alg_background}). 

The efficiency of \pbs makes it particularly suitable for solving large problem instances, or in cases where  multiple MAPF instances need to be solved rapidly. For example, \pbs has recently been used  as a building block within an approach for tackling the Lifelong MAPF (L-MAPF) problem~\cite{ma2017lifelong,Ma.Honig.ea.19,liu2019task,salzman2020research}.
In L-MAPF the agents need to execute a stream of tasks consisting of multiple locations to be visited by each agent (rather than moving to a specific goal location as in the one-shot version). A recent work~\cite{li2021lifelong} proposed an effective approach called Rolling-Horizon Collision Resolution \textsf{(\rhcr)} to solve L-MAPF, by breaking the L-MAPF problem into a \emph{sequence} of one-shot MAPF problems (see \Sec \ref{sec:alg_background}). Clearly, by improving the efficiency of the internal MAPF solver (e.g., \pbs) we can improve the efficiency of \rhcr overall, as it solves \emph{multiple} MAPF queries.

In this work we explore the use of \emph{experience} gained from solving previous MAPF queries to speed up the solution of the entire L-MAPF problem. Using experience has been considered in various domains and properly using experience to improve MAPF solvers has been identified as a key challenge and opportunity~\cite{salzman2020research}. For instance, in robot motion planning previous solutions can be reused for a new query~\cite{coleman2015experience}.
Similar ideas were employed within \cbs-based approaches to incentivize agents to traverse certain regions, e.g., corridors, in a specific manner to resolve conflicts~\cite{cohen2016bounded,li2019symmetry,li2020new}. Learning-based approaches have been used to implicitly encode experience for conflict resolution in \cbs~\cite{Huang.ea.21} and MAPF algorithm selection~\cite{Kaduri.ea.20}.
With that said, we are not familiar with systematic approaches that exploit experience within MAPF queries in L-MAPF. 

\paragraph{Contribution.}
In this paper we develop the Experienced \rhcr \textsf{(\exrhcr)} approach for L-MAPF, which allows to transfer \emph{experience} gained from solving one MAPF query to the next one in order to improve planning times (\Fig \ref{fig:timeline} and \Sec \ref{sec:algorithm}). In particular, \exrhcr solves constituent MAPF queries by interleaving between calls to the vanilla \pbs and an extension of \pbs which we call Experienced \pbs \textsf{(\expbs)} (\Fig \ref{fig:expbs_changes}). 
Unlike the standard \pbs, which begins the high-level search on a given MAPF instance from an empty priority set at the PT root, \expbs is initialized with a specific priority set for the root node that encodes the experience gained from the solution of a previous MAPF query using \pbs. This allows to reduce the depth of the search tree and speed up the solution of MAPF queries.  We also consider a lightweight version of \expbs that uses total priorities as experience, which leads to good performance in easier instances. We demonstrate that \exrhcr solves L-MAPF instances up to 39\% faster than \rhcr,
which allows to potentially increase the throughput of a given stream of task streams by increasing the number of agents we can cope with for a given time budget (\Sec \ref{sec:experiments}).

%% file: tex/probelm_definition.tex
\section{Problem Definition}\label{sec:definition}
In this section we provide a definition of  Lifelong MAPF (L-MAPF), for which we design an effective algorithmic approach in \Sec \ref{sec:algorithm}. Before describing L-MAPF, we first define the single-query, or one-shot, setting termed MAPF.

\subsection{Multi Agent Path Finding (MAPF)}
In the MAPF problem, we are given a 
graph $G = (V, E)$ and a set of $k$ agents $\calA=\{a_1, \dots, a_k\}$, where an agent $a_i\in \calA$ starts at a vertex  $s_i \in V$ and needs to reach a goal vertex $g_i \in V$. 
Time is discretized such that at each timestep, each agent occupies exactly one vertex. Between any two consecutive timesteps, each agent can either \emph{move} to an adjacent vertex (i.e., along an edge connecting its current vertex and the destination vertex) or \emph{wait} in its current vertex, where each action is assigned a unit cost.

The goal of MAPF is to find a \emph{feasible} solution plan, which consists of a set of paths $\calP = \{p_1, \dots, p_k\}$, where~$p_i$ is a path for agent $a_i$ from $s_i$ to $g_i$, such that no \emph{conflicts} arise between the different agents. 
In particular, for a given agent $a_i\in \calA$, a path $p_i=(v_{0}^i,v_{1}^i,\ldots, v_{T_i}^i)$ is a sequence of $T_i\geq 1$ vertices, such that $(v_t^i,v_{t+1}^i)\in E$ or $v_t^i=v_{t+1}^i$, and $v_0^i=s_i, v_{T_i}^i=g_i$. 
There are two types of conflicts between any two given agents $a_i\neq a_j$ to avoid: 
In a \emph{vertex conflict} the agents 
occupy the same vertex at the same timestep, i.e., $v_t^i=v_t^j$ for some $1\leq t \leq \min\{{T_i}, {T_j}\}$. 
In an \emph{edge conflict}, the agents cross the same edge at the same timestep, i.e., $v_t^i=v_{t+1}^j$ and $v_{t+1}^i=v_t^j$ 
for timestep $1\leq t < \min\{{T_i}, {T_j}\}$. 


We will also consider a relaxed version of MAPF called Windowed-MAPF (W-MAPF), which is defined for a constant parameter $w\geq 1$ indicating the window size wherein conflicts should be resolved~\cite{silver2005cooperative}. 
I.e., a solver for W-MAPF needs to avoid conflicts only up to timestep $w$, after which conflicts are allowed. This relaxation reduces the size of the search tree, in comparison to MAPF, which leads to faster computation times.
Most solvers for vanilla MAPF can be adapted to work for the windowed case. 

\subsection{Lifelong Multi Agent Path Finding (L-MAPF)}
In L-MAPF, agents need to execute a stream of tasks, where each task consists of moving between two specific vertices, while avoiding conflicts. In this setting the objective can be completing all tasks as quickly as possible, or maximizing throughput, which is the average number of tasks completed per unit of time.

A critical component within L-MAPF solvers is a \emph{task assigner}, which  specifies for each agent the next task it performs. 
There are various approaches to design task assigners~\cite{ma2017lifelong,liu2019task}, which are outside the scope of this work. 
One common approach for tackling L\nobreakdash-MAPF problems is partitioning it into a sequence of W\nobreakdash-MAPF problems, where the task assigner is invoked after every MAPF query is completed  to generate start and goal locations for the next MAPF  instance~\cite{ma2017lifelong,liu2019task,li2021lifelong}. We describe one such approach called \rhcr in the next section. 

%% file: tex/algorithmic_background.tex
\section{Algorithmic Background}\label{sec:alg_background}

\begin{figure*}[t]  
\hspace*{\fill}
    \begin{subfigure}[t]{0.32\linewidth}
        \centering
        \includegraphics[width=0.85\linewidth]{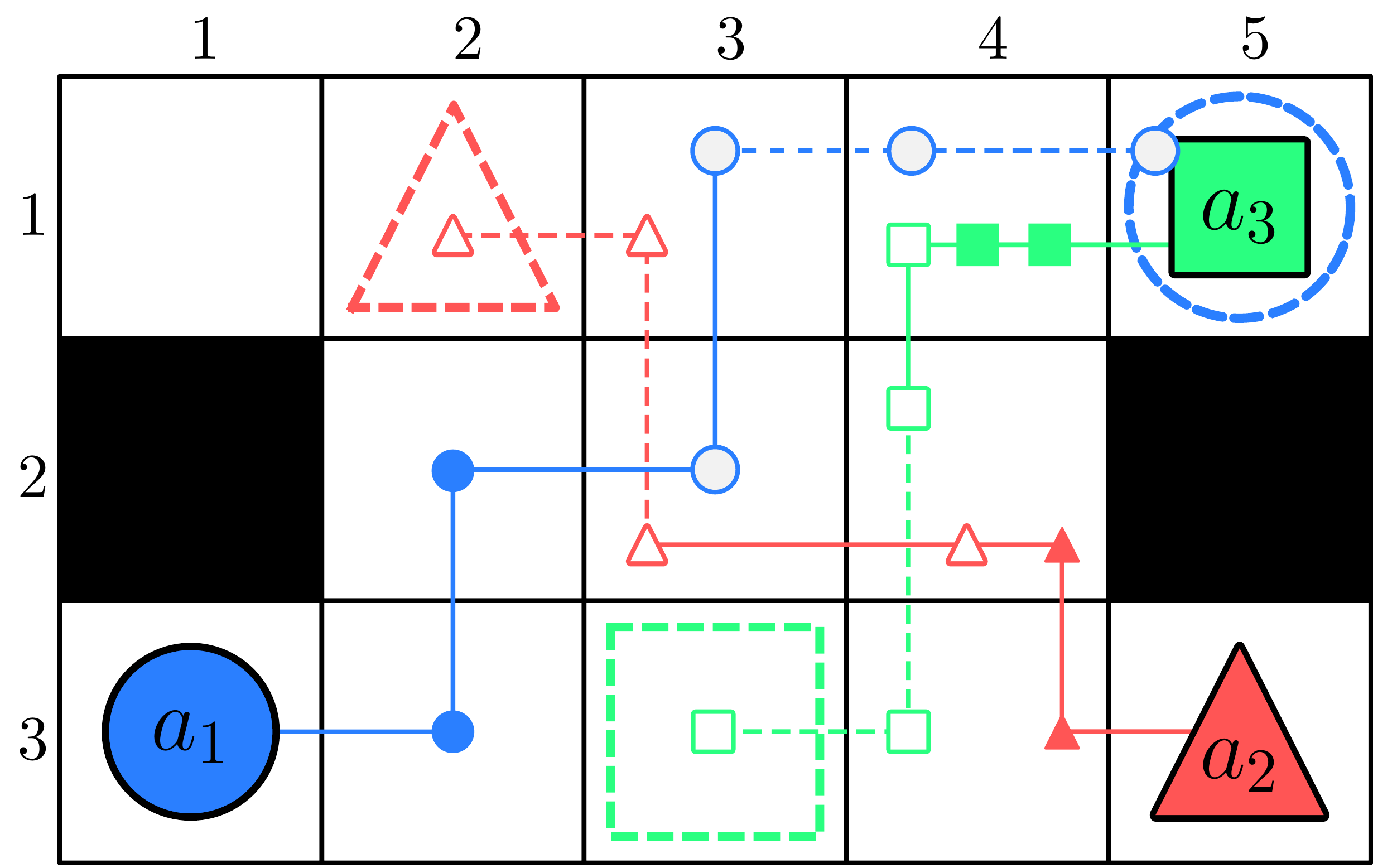}
        \caption{Initial query $q_0$ solved with \pbs. } 
        \label{fig:toy1}
    \end{subfigure}
    \hspace*{\fill}
    \begin{subfigure}[t]{0.32\linewidth}
        \centering
        \includegraphics[width=0.85\linewidth]{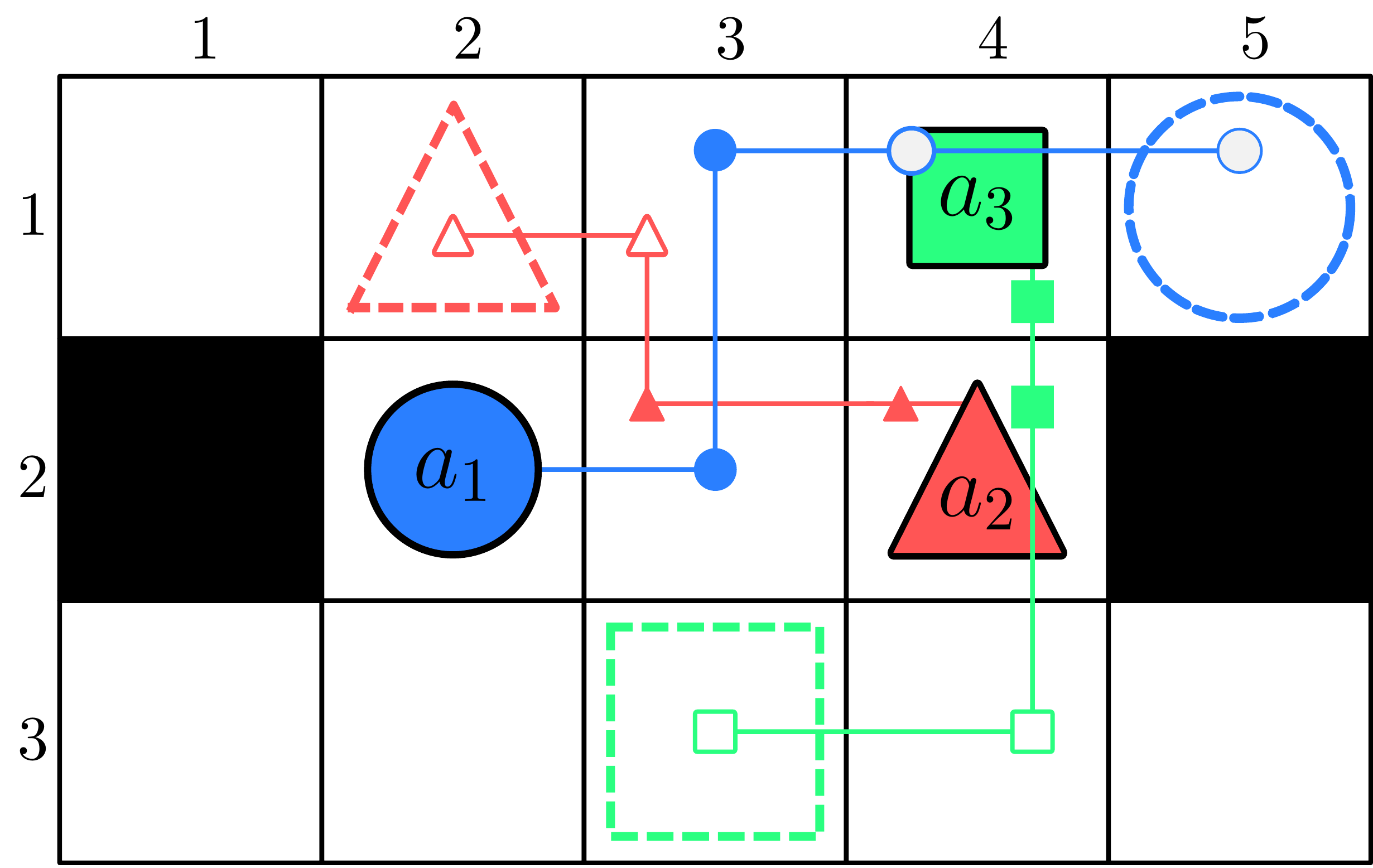}
        \caption{Query $q_1$ solved with \expbs.} 
        \label{fig:toy2}
    \end{subfigure}
    \hspace*{\fill}
    \begin{subfigure}[t]{0.32\linewidth}
        \centering
        \includegraphics[width=0.85\linewidth]{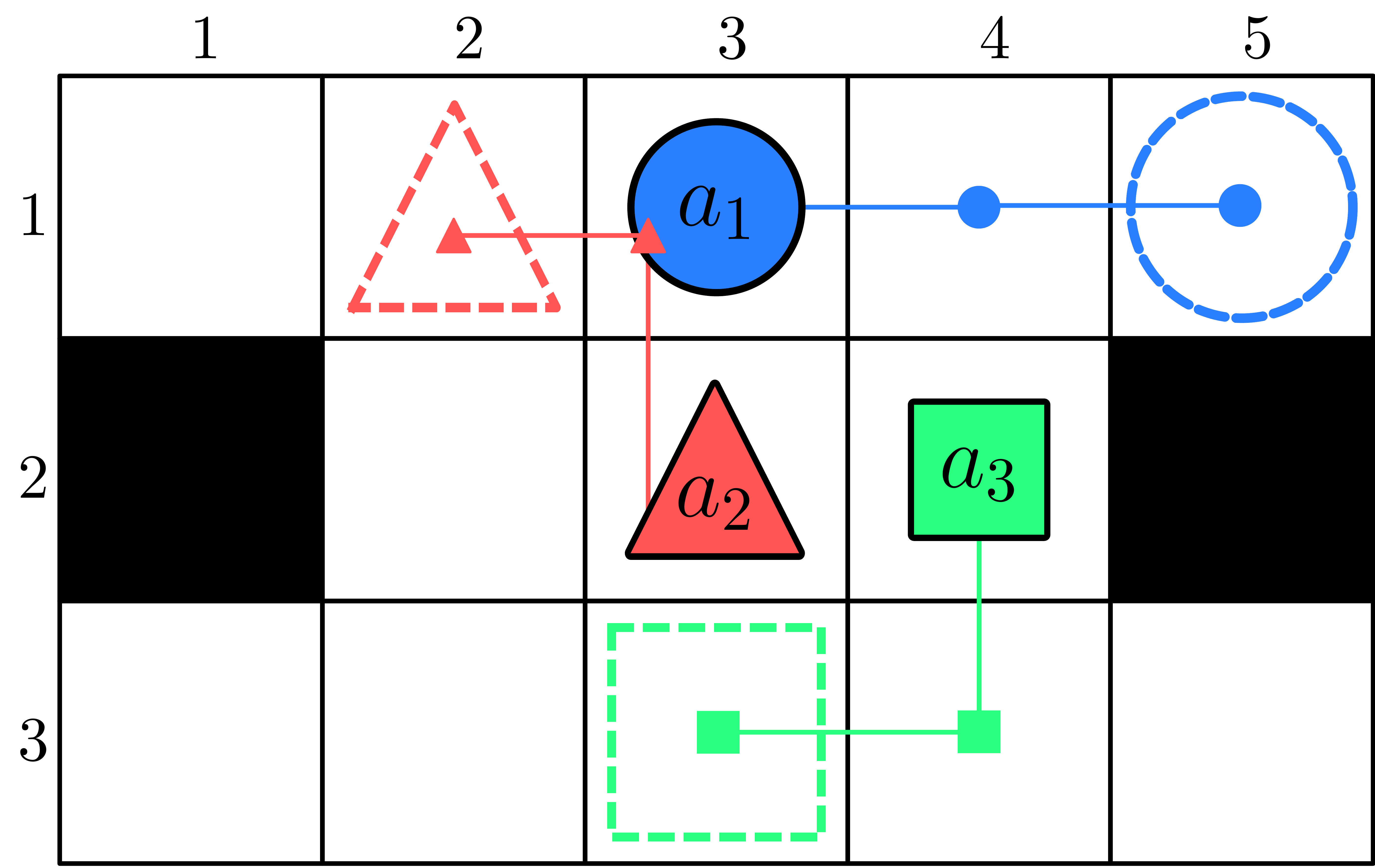}
        \caption{Query $q_2$ solved with \expbs.} 
        \label{fig:toy3}
    \end{subfigure}
    \hspace*{\fill}
    \caption{
    A toy example visualizing an execution of \rhcr and \exrhcr for three agents, whose current start and goal positions are illustrated as solid and dashed shapes, respectively, with parameters $w=4, h=2$, and $\delta= 2$ for \exrhcr. 
    Solid lines represent the path's portion up to the $w$ first timesteps (where collisions are avoided) and dashed line represent the rest (where collisions are ignored). 
    Solid markers on the paths represent the first $h$ steps and the rest are represented by hollow markers. Here, we chose \expbs and \pbs to break ties in the low and high-level search in a specific manner, but other valid execution outcomes are possible.
    (\protect\subref{fig:toy1}) The W-MAPF instance $q_0$ with its solution obtained by \pbs, where the experience gained is \PRECexp $ = \{ a_2 \prec a_3, a_1 \prec a_2\}$, which forces $a_2$ to avoid $a_1$'s path, and for $a_3$ to avoid the paths of $a_1$  and $a_2$  (which causes~$a_3$  to wait). See full details of \pbs's execution in the \myappendix. 
    (\protect\subref{fig:toy2}) and (\protect\subref{fig:toy3}) show the next two queries~$q_1$ and~$q_2$ where they both use  $\expbs(\PREC_{\text{exp}})$, which allows to obtain a feasible solution already in the tree root, i.e., without further expansion of PT nodes.
    In contrast, to solve $q_1$ \rhcr would expand two PT nodes representing the same conflicts in $q_0$ (to see this, note that for some shortest paths the agents do collide at the same positions as in $q_0$).
    In both (\protect\subref{fig:toy2}) and (\protect\subref{fig:toy3}) \exrhcr finds the continuation of the solution found in (\protect\subref{fig:toy1}). 
    In (\protect\subref{fig:toy2}) $a_2$ waits in place to avoid $a_1$'s path, and $a_3$ waits in its start position to avoid $a_2$. 
    In (\protect\subref{fig:toy3}), query~$q_2$, the agents can follow their shortest paths without colliding, thus \rhcr finds a solution in the first PT node without expanding further nodes. \exrhcr also finds a solution from the first node although an experience is given. 
    }
    \label{fig:toy_example}
\end{figure*}

We describe two algorithmic components, which we will build upon to design our approach for L-MAPF (\Sec \ref{sec:algorithm}). In particular, we describe Priority-Based Search \textsf{(\pbs)} for MAPF, and  Rolling-Horizon Collision Resolution \textsf{(\rhcr)} for L-MAPF.

\subsection{Priority-Based Search \textsf{(\pbs)}} \label{sec:pbs}

\pbs is a recent approach for MAPF that can be used to solve  instances with a large number of agents, or situations where multiple MAPF instances need to be solved rapidly (as in L-MAPF).
We now provide an overview of \pbs, and refer the reader to~\cite{ma2019searching} for the full description. 

\pbs 
maintains priorities between agents which are used to resolve conflicts.
On the high-level search \pbs explores a \emph{priority tree} (PT), where a given node~$N$ of PT encodes a (partial) priority set $\PREC_N=\{a_h \prec a_i, a_j \prec a_l$, \ldots \}. A priority $a_{i}\prec a_{j}$ means that agent $a_{i}$ has precedence over agent~$a_{j}$ whenever a low-level search is invoked (see below). In addition to the ordering, each PT node maintains single-agent plans that represent the current MAPF solution (possibly containing conflicts).
\pbs starts the high-level search with the tree root whose priority set is empty, and assigns to each agent its shortest path. We mention that \pbs  can be initialized with a non-empty priority set in its root node, albeit previous work has not specified an effective method to specify this set.
Whenever \pbs expands a node $N$, it invokes a low-level search to compute a new set of plans which abide to the priority set $\PREC_N$. If a collision between agents, e.g., $a_i$ and $a_j$, is encountered in the new plans, \pbs generates two child PT nodes $N_1, N_2$ with the updated priority sets $\PREC_{N_1}=\PREC_N\cup \{a_i\prec a_j\}, \PREC_{N_2}=\PREC_N\cup \{a_j\prec a_i\}$, respectively. The high-level search chooses to expand at each step a PT node in a depth-first search \textsf{(\dfs)} manner. 
The high-level search terminates when a valid solution is a found at some node $N$, or when no more nodes for expansion remain, in which case \pbs declares failure. 

The low-level search of \pbs proceeds in the following manner. For a given PT node $N$, \pbs performs a topological sort of the agents according to $\PREC_N$ from high priority to low, and plans individual-agent paths based on the sorting. 
Given a topological sort $(a'_{1},\ldots, a'_{k'})\subset \calA$, for some $1\leq k'\leq k$, the low-level search iterates over the~$k'$ agents in the topological sort, and updates their plans such that they do not collide with any higher-priority agents 
(agents that do not appear on this list maintain their original plans). It then checks whether collisions occur between all the agents combined.

\subsection{Rolling-Horizon Collision Resolution \textsf{(\rhcr)}}\label{sec:rhcr}

\rhcr~\cite{li2021lifelong}, is a state-of-the-art framework for solving L-MAPF. \rhcr accepts as parameters a time window $w\geq 1$, and a replanning rate $h\leq w$, and decomposes the L-MAPF  problem into a sequence of  W-MAPF queries $q_0, q_1, \ldots$ that are solved one by one. In particular, after obtaining a W-MAPF plan for a query  $q_i$ with time window $w$, the next query $q_{i+1}$ is generated by executing the solution plan for $q_{i}$ for $h$ steps. This yields the agent's start locations for query $q_{i+1}$. The goal locations remain the same for agents that did not complete their task and new goal locations are assigned by the task assigner for agents that did.   
For a concrete example of a task assigner, see \Sec \ref{sec:experiments}.

\rhcr requires a W-MAPF solver as an internal submodule. 
Accordingly, several bounded-horizon versions of state-of-the-art MAPF algorithms were used, among which \pbs proved to be the most effective as a W-MAPF solver within \rhcr.  Overall, it was observed that by using a small time window~$w$, and consequently a small replaning rate~$h$, \rhcr can obtain faster solutions than alternative approaches. This, in turn, allows to solve queries containing more agents and potentially improve the system's throughput~\cite{li2021lifelong}. 



%% file: tex/algorithm.tex
\section{Leveraging Experience in Lifelong MAPF}\label{sec:algorithm}

In this section we present our algorithmic approach for leveraging experience in L-MAPF, which we call Experienced \rhcr \textsf{(\exrhcr)}. First we discuss some properties of the original \rhcr framework, which will be instrumental in the development of \exrhcr. 

Clearly, by improving the efficiency of the internal MAPF solver (e.g., \pbs) we can improve the efficiency of \rhcr overall, as it solves \emph{multiple} W-MAPF queries. However, speeding up MAPF solvers without leveraging additional structure within the \rhcr approach can be difficult. Next, we observe that there is an underlying structure emerging from the fact that the W-MAPF queries are {\em consecutive}, i.e., a solution plan to one query $q_i$ defines the next query $q_{i \text{+}1}$. In particular,  considering that a short replan rate $h$ is typically used by \rhcr, consecutive W-MAPF queries can be quite similar to one another in terms of initial and goal agent locations. This, in turn, can lead to \emph{similar} queries, in terms of the conflicts that need to be avoided and the agents between which they arise. Despite this, current \rhcr implementations solve every subsequent W-MAPF query \emph{from scratch}. 

In contrast to \rhcr, our approach allows to transfer \emph{experience} gained from solving one W-MAPF query to the next one in order to improve planning times. In particular, \exrhcr solves W-MAPF queries by interleaving between calls to the vanilla \pbs and an extension of \pbs which we call Experienced \pbs \textsf{(\expbs)}. Unlike the standard \pbs, which begins the high-level search on a given W-MAPF instance from an empty priority set~$\emptyset$ at the root of the PT, \expbs is initialized with a specific priority set~$\PREC_\text{exp}$ for the root of the PT. This priority~$\PREC_\text{exp}$ encodes the experience gained from the solution of a previous W-MAPF query obtained via \pbs. See illustration of \exrhcr and \expbs in \Fig \ref{fig:page1figure}. Additionally, in \Fig \ref{fig:toy_example} we illustrate two consecutive \wmapf sub-queries ($q_0$ and $q_1$) within an L-MAPF problem, that have the same conflicts in terms of types of agents and positions in case that \pbs is used for solving both queries. However, the conflicts in $q_1$ can be avoided by reusing the experience obtained in $q_0$ to solve $q_1$ using \expbs.

Encoding experience as priorities has the desirable property of being {\em generalizable} in the sense that a priority  used for solving one  query can be applied to a similar query  without necessarily over-constraining the solution. This is because priorities provide high-level specification that are likely to apply in a variety of queries, rather than hard constraints which need to be followed exactly (e.g., the constraints used in \cbs that specify for each agent exact locations that should be avoided in specific points in time). Experience encoded in the form of a 
priority set can thus be thought as a way to warm-start consecutive MAPF queries.

In some cases previous experience might not be useful for the given query, in which case we want to ensure the {\em robustness} of our MAPF solver to uninformative experience. In \expbs this is achieved by restarting the search with an empty priority set in case that experience causes the search to diverge. In the remainder of this section we detail  \exrhcr and the internal MAPF solver \expbs.

\subsection{Experienced \rhcr \textsf{(\exrhcr)}}\label{sec:exrhcr}

The \exrhcr algorithm accepts as inputs a graph $G$ representing the environment, and a task assigner $\calT$, which implicitly maintains the agent's current start and goal positions as well as the remaining tasks. Using the task assigner as a black box helps keeping our description below generic. Similarly to \rhcr, \exrhcr has the parameters of time window~$w$, and replanning rate~$h$. Two additional parameters that we introduce for \exrhcr are the \emph{experience lookahead}~$\delta$, which determines the number of \expbs calls after every \pbs call (see details below), and the \expbs high-level  PT \emph{width limit}~$\ell> 1$. 
The latter parameter helps to identify situations where previous experience turns out to be uninformative for the current query, and to restart the high-level search with an empty priority set (see \Sec \ref{sec:expbs}).



The motivation behind using the lookahead parameter $\delta$ is keeping the experience up-to-date, and employing it only when it is likely to be relevant to the current query,  which should not differ vastly from the query from which the experience was extracted. 
We also ensure the experience does not become overconstrained by generating a brand new experience every few runs (rather than, e.g., constantly passing the priority set obtained from solving the current query as experience to the next query, and so on, where  priorities are accumulated from each run).

%
We suggest to set $\delta$ around the value $\lfloor \frac{w}{h}\rfloor-1$, which means the experience is used in all the queries of the planning horizon it was created in
to maximize the experience utilization.
See more details in \Sec \ref{sec:exp_delta}.

\exrhcr is detailed in \Alg \ref{alg:exrhcr} with the differences from \rhcr highlighted in blue. 
\exrhcr first uses the task assigner to obtain the initial W-MAPF instance~$q_0$ and initializes to zero the counter $i$ which represent the current \wmapf query index [Line~\ref{exrhcr:initial_state}]. 
It then iteratively solves~$q_i$ and all subsequent W-MAPF instances until all tasks are completed [Line~\ref{exrhcr:while}].
It runs \pbs with the time window~$w$ on query~$q_i$ to obtain a plan~$\calP_{q_i}$, which specifies agent paths for the current W-MAPF instance~[Line~\ref{exrhcr:pbs_run}]. 
In contrast to \rhcr, the priority set of the PT node for which~$\calP_{q_i}$ was obtained is stored in memory. We term this the \emph{seed priority set} and denote it by $\PREC_\text{exp}$. 
Subsequently, and similar to \rhcr, a new W-MAPF query~$q_{i\text{+}1}$ is generated by the task assigner, and the counter~$i$ is updated~[Line~\ref{exrhcr:TA_pbs}]. This is done by executing the plan $\calP_{q_i}$ for $h$ steps to update the agents' start locations, and potentially updating the agents' goals and tasks. 
At this point, \exrhcr differs from \rhcr:
for the next~$\delta$ W-MAPF instances [Line~\ref{exrhcr:for_delta}], unless all tasks are finished [Line~\ref{exrhcr:delta_ta_check}], the algorithm invokes \expbs with the seed priority set $\PREC_\text{exp}$ (using the same window size $w$ and with a parameter $\ell$, which will be explained shortly) [Line~\ref{exrhcr:expbs_run}].
Given the new solution plan $\calP_{q_i}$ and replaning rate $h$, the query and the counter are updated as before [Line~\ref{exrhcr:TA_expbs}] and this inner loop is repeated.

\begin{algorithm}[t]
\caption{\textcolor{blue}{\textsf{ex}}\rhcr}
\label{alg:exrhcr}
\textbf{Inputs:} L-MAPF query, graph $G$, task assigner $\calT$ 
\label{exrhcr:inputs} \\
\textbf{Parameters:} Window size $w$, replanning rate $h$, \textcolor{blue}{experience lookahead $\delta$, width limit $ \ell$} \label{exrhcr:params} \\
\textbf{Output:} Paths for all agents\label{exrhcr:output}
\begin{algorithmic}[1] 
\STATE $q_0 \gets\calT.\textsc{InitialQuery}()$; \ \ $i=0$; \label{exrhcr:initial_state} 

\WHILE{not $\calT.\textsc{Empty}()$}\label{exrhcr:while}
    \STATE $(\calP_{q_{i}}, \textcolor{blue}{\PREC_\text{exp}}) \gets$ \pbs($q_i, w$); \label{exrhcr:pbs_run}
    \STATE $q_{i\text{+}1} \gets \calT.\textsc{NextQuery}(q_{i}, \calP_{q_{i}}, h)$; \ \  $i\text{++}$; \label{exrhcr:TA_pbs}
    
    \begingroup \color{blue}
    \STATE \textbf{repeat} $\delta$ \textbf{times}\label{exrhcr:for_delta}
    \COMMENT{run \expbs with \pbs as fallback}
        \STATE \ \ \ \ \textbf{if} $\calT.\textsc{Empty}()$ \textbf{return}\label{exrhcr:delta_ta_check} \COMMENT{tasks finished}
        \STATE \ \ \ \ $\calP_{q_i} \gets$ \expbs($q_i,\PREC_\text{exp},w,\ell$);\label{exrhcr:expbs_run}
        \STATE \ \ \ \ $q_{i\text{+}1}\gets \calT.\textsc{NextQuery}(q, \calP_{q_{i}}, h)$; \ \  $i\text{++}$;\label{exrhcr:TA_expbs}
    \endgroup
\ENDWHILE

\end{algorithmic}
\end{algorithm}

\subsection{Experienced \pbs \textsf{(\expbs)}} \label{sec:expbs}

\expbs utilizes \pbs's option of starting the high-level search from a given seed priority set which we denote as an experience rather than from the typically-used empty priority set.
An additional difference is that \expbs avoids overexploring the PT by limiting the width of the explored PT rooted in $\PREC_\text{exp}$, in case that~$\PREC_\text{exp}$ does not lead to a solution fast enough (or does not find a solution at all), and restarts the search with a priority $\emptyset$ by calling the ``vanilla'' \pbs. We call the usage of \pbs after \expbs terminates with no solution the ``fallback''.
To limit the tree width, \expbs explores the high-level PT using a \emph{Width-Limited Depth-First Search} \textsf{(\wldfs)}, which we describe below, rather than the standard \dfs used by \pbs. Thus, \expbs accepts two additional parameters when compared  to \pbs: the seed priority set~$\PREC_\text{exp}$ and the width-limit parameter $\ell$.

We provide additional details on \wldfs. Given a tree graph, let its \emph{width} denote the maximal number of nodes across all levels (where two nodes are on the same tree level if their distance from the root, or depth, is the same). 
The parameter $\ell$ specifies the maximal width allowed when exploring a PT using \wldfs. To keep track of the current width of the PT we maintain for each level a counter representing the number of nodes in the level, and increment it whenever new nodes are added. 
When the width of the PT exceeds $\ell$ (or when no solution exists), \wldfs aborts the search of the PT rooted in $\PREC_\text{exp}$, and invokes the vanilla \pbs solver without width limitation. 

We chose to limit the search efforts by limiting the PT width as we found that it is a good indicator whether the experience is over-constrained or not (over-constrained  experience tend to force the search to expand entire, wide subtrees).
We found (see \Sec  \ref{sec:experiments}) that alternative measures such as the number of expanded nodes require per-instance tuning as \expbs uses a Depth First Search (\dfs) on the PT and thus we need to account for the scenario
and the number of agents. In contrast, we empirically found that the width is a robust parameter that does not require tuning. Indeed, in all our experiments (\Sec  \ref{sec:experiments}) we used the same width value.
Finally, in \Sec \ref{sec:exp_ell} we study the effect of \wldfs with different values of $\ell$ on the overall performance and show the robustness of the method to the specific choice of $\ell$.

\subsection{Alternative Instantiation with Total Priority}
\label{sec:alt}
\begin{figure}[t!]  
    \begin{subfigure}[t]{\columnwidth}
        \centering
        \includegraphics[height=0.48\linewidth]{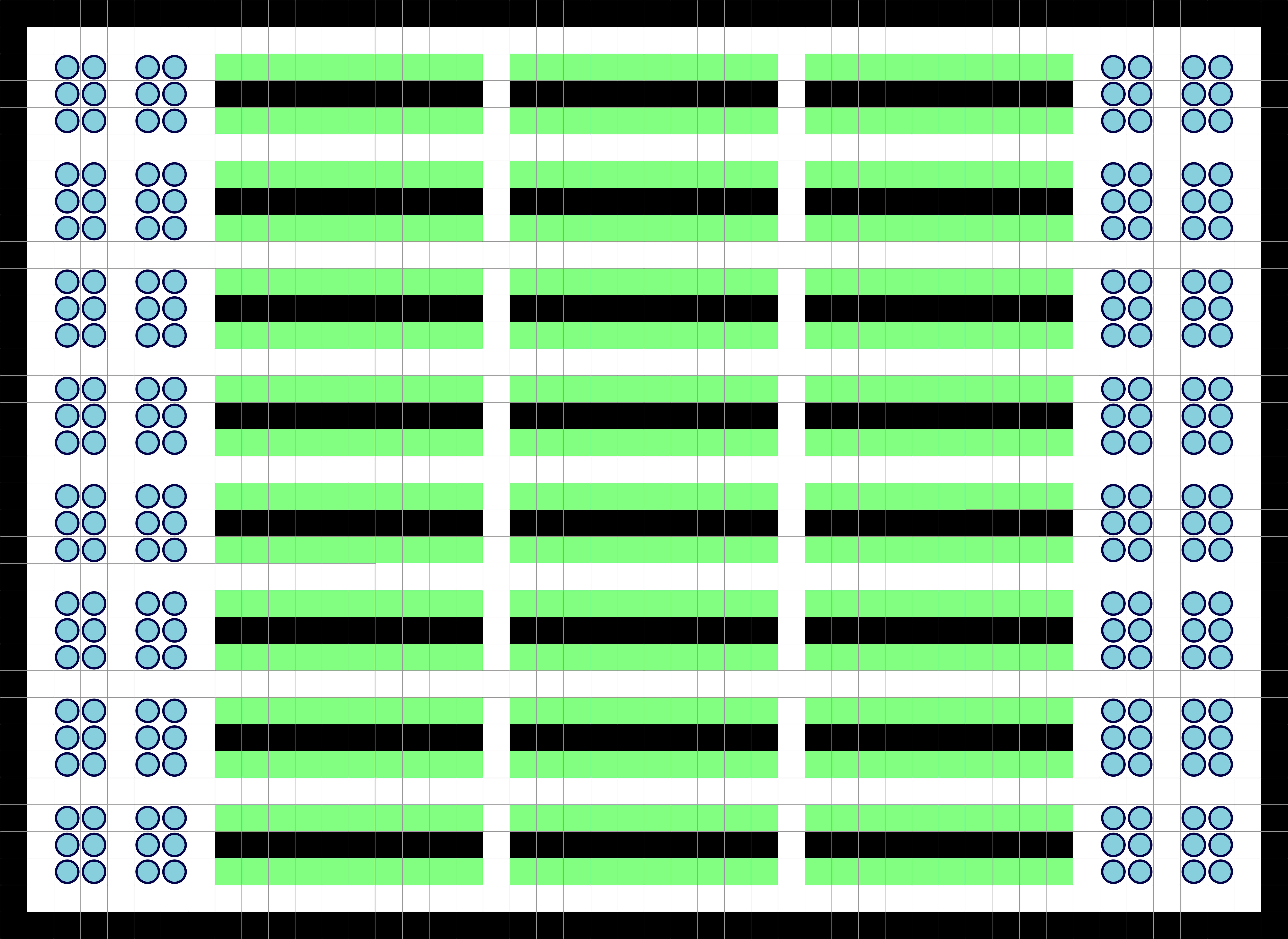} 
        \caption{\warehouse environment.}
        \label{fig:scenario_kiva}
    \end{subfigure}
    \hspace{15mm}
    \begin{subfigure}[t]{\columnwidth}
        \centering
        \includegraphics[height=0.48\linewidth]{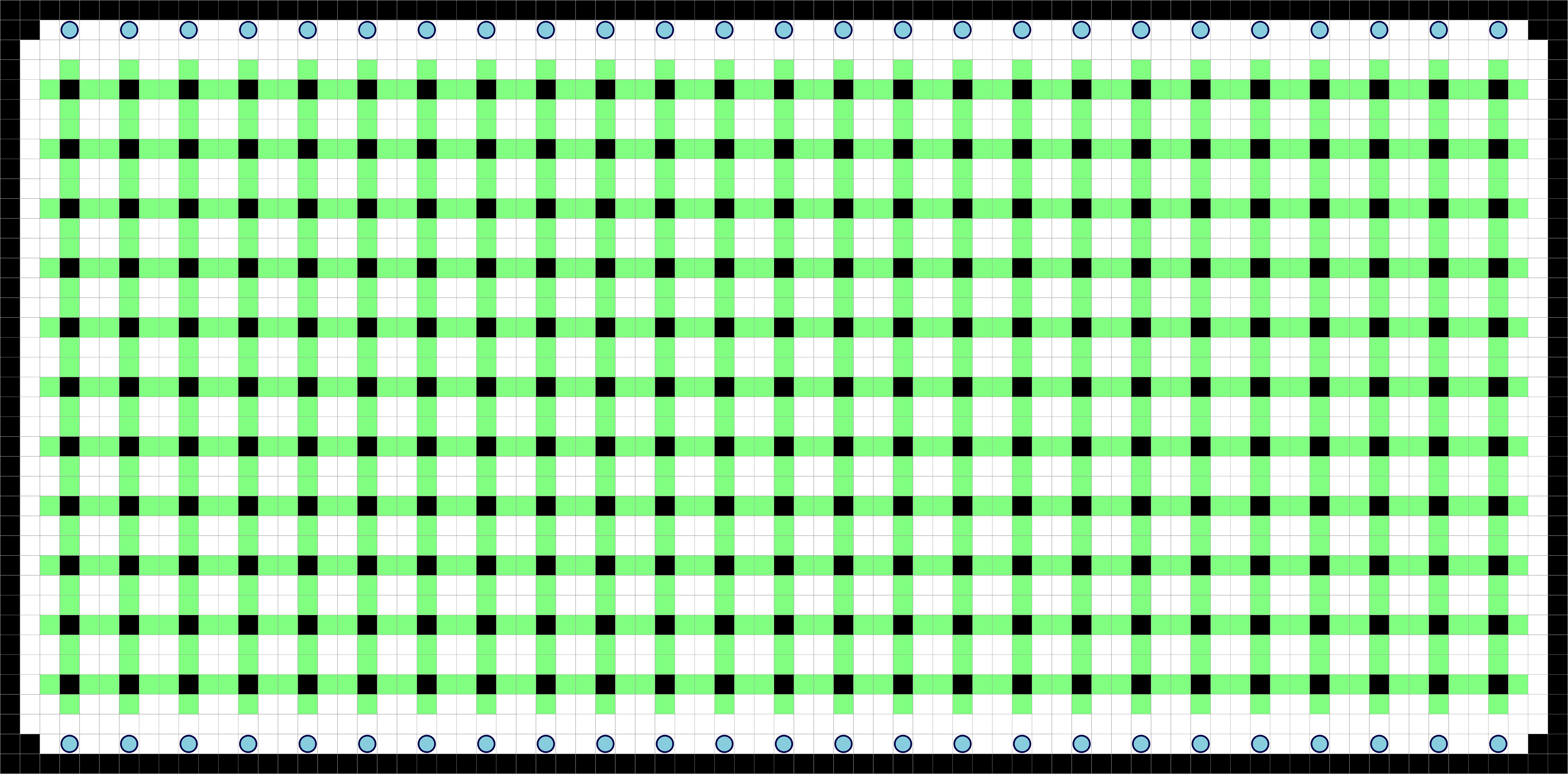} 
        \caption{\sorting environment.}
        \label{fig:scenario_sorting}
    \end{subfigure}
    \hspace*{\fill}
    \caption{Benchmark environments for L-MAPF problems.
    \protect{(\subref{fig:scenario_kiva})}~A $33 \times 46$ \warehouse domain with ${\sim} 16\%$ obstacles (black) representing $240$ pods~\protect\cite{liu2019task}.
    Working stations are drawn in blue and task locations around the inventory pods are drawn in green. 
    \protect{(\subref{fig:scenario_sorting})}~A $37 \times 77$ \sorting center with ${\sim} 10\%$ obstacles representing chutes~\protect\cite{li2021lifelong}.  
    Working stations are marked in blue, and task locations, which represent drop-off locations around chutes, are shown in green.
    Note that we used the undirected versions of these benchmarks.
    }
    \label{fig:benchmarks}
\end{figure}


\begin{figure*}[t]  
    \centering
    \hspace*{\fill}
    \begin{subfigure}[t]{0.95\columnwidth}
        \centering
        \includegraphics[width=\columnwidth]{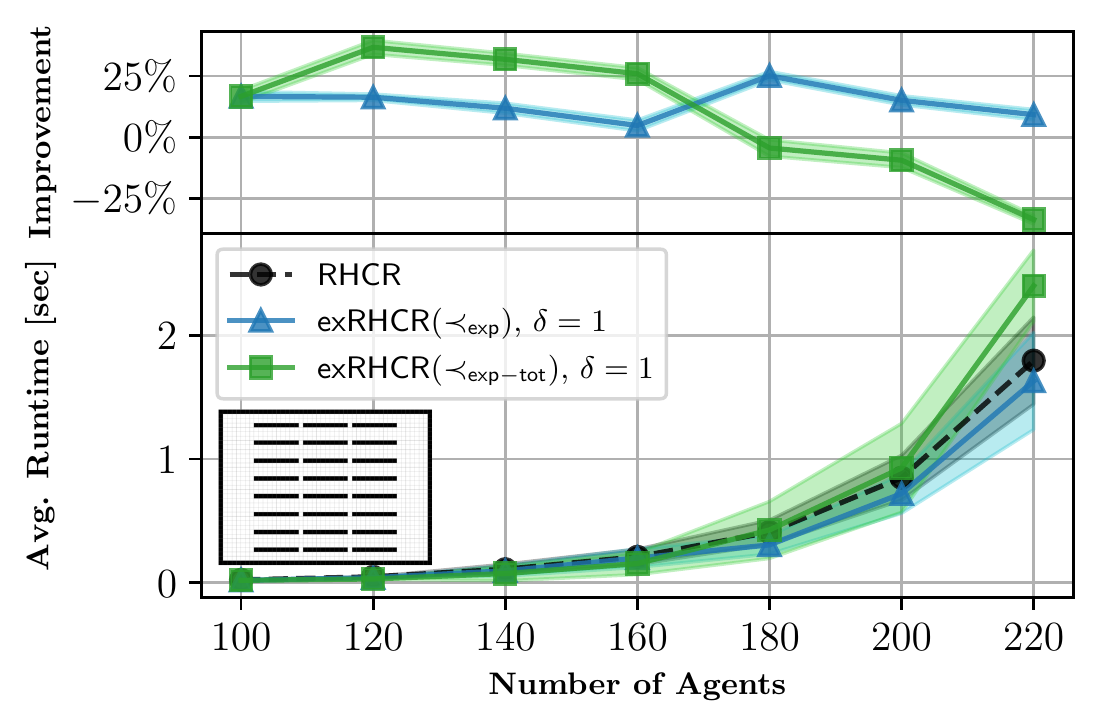}
        \caption{Runtime for \warehouse.}
        \label{fig:lifelong_kiva}
    \end{subfigure}
    \hspace*{\fill}
    \begin{subfigure}[t]{0.95\columnwidth}
        \centering
        \includegraphics[width=\columnwidth]{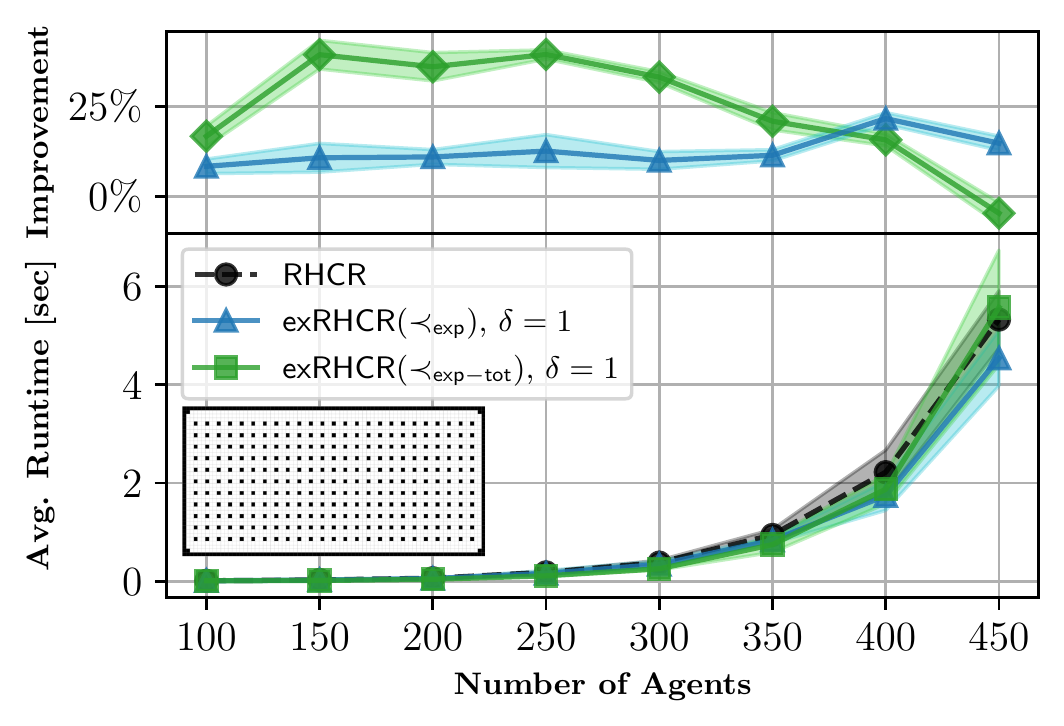} 
        \caption{Runtime for \sorting.} 
        \label{fig:lifelong_sorting}
    \end{subfigure}
        \hspace*{\fill}
        \\  
        \hspace*{\fill}
    \begin{subfigure}[t]{0.95\columnwidth}
        \centering
        \includegraphics[width=\columnwidth]{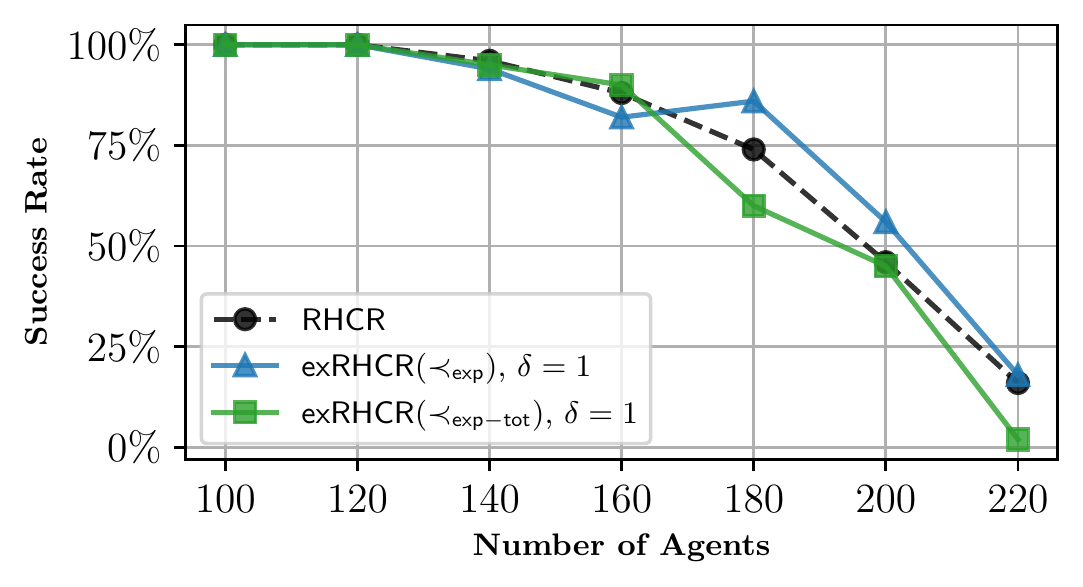} 
        \caption{Success Rate for \warehouse. }
        \label{fig:lifelong_success_rate_kiva}
    \end{subfigure}
    \hspace*{\fill}
    \begin{subfigure}[t]{0.95\columnwidth}
        \centering
        \includegraphics[width=\columnwidth]{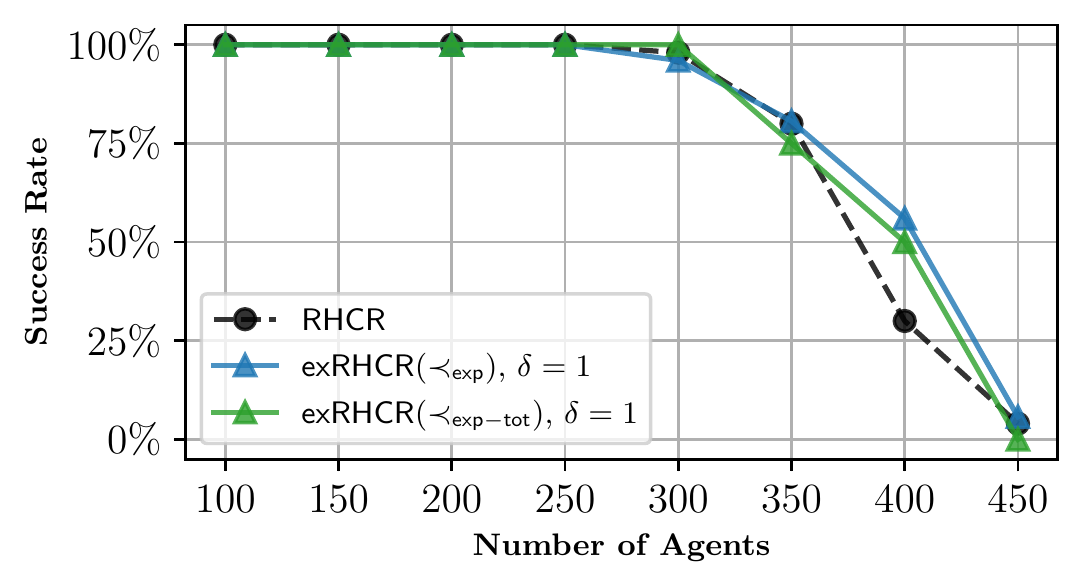} 
        \caption{Success Rate for \sorting. }
        \label{fig:lifelong_success_rate_sorting}
    \end{subfigure}
    \hspace*{\fill}
    \caption{
    L-MAPF experiments results.
    \protect{(\subref{fig:lifelong_kiva})} and
    \protect{(\subref{fig:lifelong_sorting})} depict the average runtimes and standard deviation of a W-MAPF query for the different L-MAPF solvers.
    For each environment, the top plot shows the relative improvement of each method  compared to~\rhcr.
    \protect{(\subref{fig:lifelong_success_rate_kiva})} and
    \protect{(\subref{fig:lifelong_success_rate_sorting})} depict the success rate obtained by all planners.
    }
    \label{fig:lifelong}
\end{figure*}
The key ingredients of our algorithmic framework are
(i)~the notion of experience derived from a non-informed planner (partial priority and \pbs in the method described)
and
(ii)~how the experience is used in the W-MAPF planner (\expbs in the method described).
Here, we suggest an alternative instantiation and discuss its merits.

Recall that after running \pbs, 
we obtain a partial priority $\PREC_\text{exp}$, which is used within \expbs. Here we introduce a lightweight alternative which computes a \emph{total priority}\footnote{A total priority specifies priorities between all the agents of the form $a'_1\prec a'_2\prec \ldots \prec a'_n$.}~$\PREC_\text{exp-tot}$ that is consistent with $\PREC_\text{exp}$ (namely, if $a_i \prec a_j$ in $\PREC_\text{exp}$ then $a_i \prec a_j$ in $\PREC_\text{exp-tot}$).  Note that such a consistent total priority always exists and is easy to compute.
Now, we run \expbs with the seed $\PREC_\text{exp-tot}$, which boils down to running a prioritized planner~\cite{silver2005cooperative} and running \pbs in case of failure.  

Using a total priority is less generalizable than a partial priority and thus the planner is more likely to fall back to \pbs. On the other hand, running a prioritized planner is extremely fast and when the problem is ``easy'' it may often succeed even with this more constrained notion of experience. 
As 
we will see in our experiments, this approach is advantageous in easier settings due to its simplicity.

%% file: tex/empirical_results.tex
\section{Experimental Results}
\label{sec:experiments}

We provide an empirical evaluation of our \exrhcr approach and compare it to \rhcr. 
We implemented the algorithms in \Cpp and tested them on an Ubuntu machine with \SI{4}{GB} RAM and a \SI{2.7}{GHz} Intel i7 CPU. We used benchmarks simulating a warehouse (\Fig \ref{fig:scenario_kiva}) and a sorting center (\Fig \ref{fig:scenario_sorting})\footnote{Code and benchmarks are available at \href{https://github.com/NitzanMadar/exPBS-exRHCR}{https://github.com/NitzanMadar/exPBS-exRHCR}. 
}.

For both environments 
we randomly initialize the start location for each agent from all possible locations, and the task assigner samples a goal location randomly from the blue and green locations depicted in \Fig \ref{fig:benchmarks}.
Each time an agent reaches a goal location of a specific color the task assigner specifies a new goal location uniformly at random from the other color that is not currently assigned to another agent.

In Section~\ref{sec:exp_lmapf} we compare our \exrhcr framework with \rhcr.
Next, in Sections~\ref{sec:exp_delta} and~\ref{sec:exp_ell} we focus on \exrhcr with partial priorities, which has more promise in tackling hard L-MAPF instances, and discuss the effect that the lookahead~$\delta$ and the width limit $\ell$ parameters have on \exrhcr and \expbs, respectively. 


\subsection{L-MAPF Experiments}
\label{sec:exp_lmapf}

\ignore{
The premise of our work is that when the replanning rate $h$ is smaller than the time window $w$, the potential collisions for the latter $w-h$ timesteps of the previous W-MAPF
instance may be avoided for the current W-MAPF instance by inheriting the priority set from the previous W-MAPF solution. 
Note that setting any total order on all agents as the initial priority set to \pbs could also potentially prevent those collisions.
To this end, as a second baseline, we compare our approach to an \exrhcr-like algorithm which uses a total priority as an ``experience'' and uses \pbs as fallback (similar to \expbs).
We denote this baseline as \rhcr$(\PREC_{\text{tot}}\rightarrow\pbs)$. \KS{Explain how $\PREC_{\text{tot}}$ is generated. }
}


For each of the two environments (i.e., \warehouse and \sorting) we generated multiple L-MAPF instances and tested them on a varying numbers of agents $k$. 
In particular, we set $k\in \{100,120,\ldots,220\}$ and ${k\in\{100,150,\ldots,450\}}$ for \warehouse and \sorting, respectively. For each combination of an environment and~$k$, we randomly generated~$50$ L-MAPF instances. 


We consider three planners, \rhcr, and our two planners: (i)
using a partial priority with \expbs, denoted by  $\exrhcr(\PREC_{\text{exp}})$ 
and (ii) the alternative instantiation described in Section~\ref{sec:alt} 
using a total priority with prioritized planning, denoted by  $\exrhcr(\PREC_{\text{exp-tot}})$.
We set the replanning rate~$h$ and the time window $w$ to be 
$h=5$ and $w=10$ for both \warehouse
and \sorting environments.
For \exrhcr we used an experience lookahead of~$\delta = \lfloor \frac{w}{h} \rfloor -1 = 1$, and width limit $\ell=10$ when using partial priority experience.

In \Fig \ref{fig:lifelong}, we report the solvers' average MAPF query runtime for a total L-MAPF execution of $250$ timesteps. 
In particular, each of the $50$ L-MAPF instances induces $50$ W-MAPF instances (a total of $250$ timesteps divided by the replan rate of $h=5$). 
We report the average runtime and standard deviation across all W-MAPF instances solved, as well as success rates.
We limit the runtime for a W\nobreakdash-MAPF query to $30$ seconds, after which we declare failure. In such a case the runtime of a failed W-MAPF instance is $30$ seconds.  




\exrhcr  with $\delta = 1$ improved the average runtime (over \rhcr) up to~$37\%$ in  \warehouse and~$39\%$ in \sorting. 
\ignore{  
Interestingly, using $\delta=2$ in \warehouse enabled an improvement between $8\%$ and $33\%$. 
For a deeper analysis on the effect of $\delta$, see \Sec \ref{sec:exp_delta}. 
}
When considering ${\exrhcr(\PREC_{\text{exp-tot}})}$, we can see (as in \citet{ma2019searching}) that the approach is highly effective when the number of agents is small. However, given a large number of agents, this method has a high failure rate (fallback is used roughly $86\%$ of the time for the largest number of agents). For a large number of agents ${\exrhcr(\PREC_{\text{exp}})}$  achieves the best performance compared to \rhcr and ${\exrhcr(\PREC_{\text{exp-tot}})}$.

\ignore{
\NM{In addition, it shown that total priority ordering (with \pbs as fallback) work better than \exrhcr as the number of agent is smaller. But, its performance is dramatically decrease as $k$ increases, and the average runtime was higher than all both \rhcr and \exrhcr.
This phenomenon is also shown in \cite{ma2019searching} when testing total priority ordering techniques compared to \quot{vanilla} \pbs.
This can be explained due to the success rate decrease and the number of fallback used ($\sim 70 \%$ when using the maximal $k$)}
}


We provide a few more observations. 
(i)~In all the {L\nobreakdash-MAPF} experiments we performed, the difference in {the average} solution cost between {\pbs and \expbs} was negligible (roughly $\pm1\%$), which suggests that reusing experience does not hinder solution quality. 
(ii)~As a result, the throughput difference per instance and number of agents is also negligible (roughly $\pm2\%$).
(iii)~The improved runtime we described above (\Fig \ref{fig:lifelong}) suggests that for a given time budget and an average W-MAPF query, \exrhcr can accommodate more agents than \rhcr. This suggests that the improved efficiency of our approach can improve the overall throughput in automated logistic domains.

\subsection{Effect of Lookahead Parameter $\delta$  on \rhcr}
\label{sec:exp_delta}

\begin{figure}[t]  
    \centering
    \includegraphics[width=0.95\linewidth]{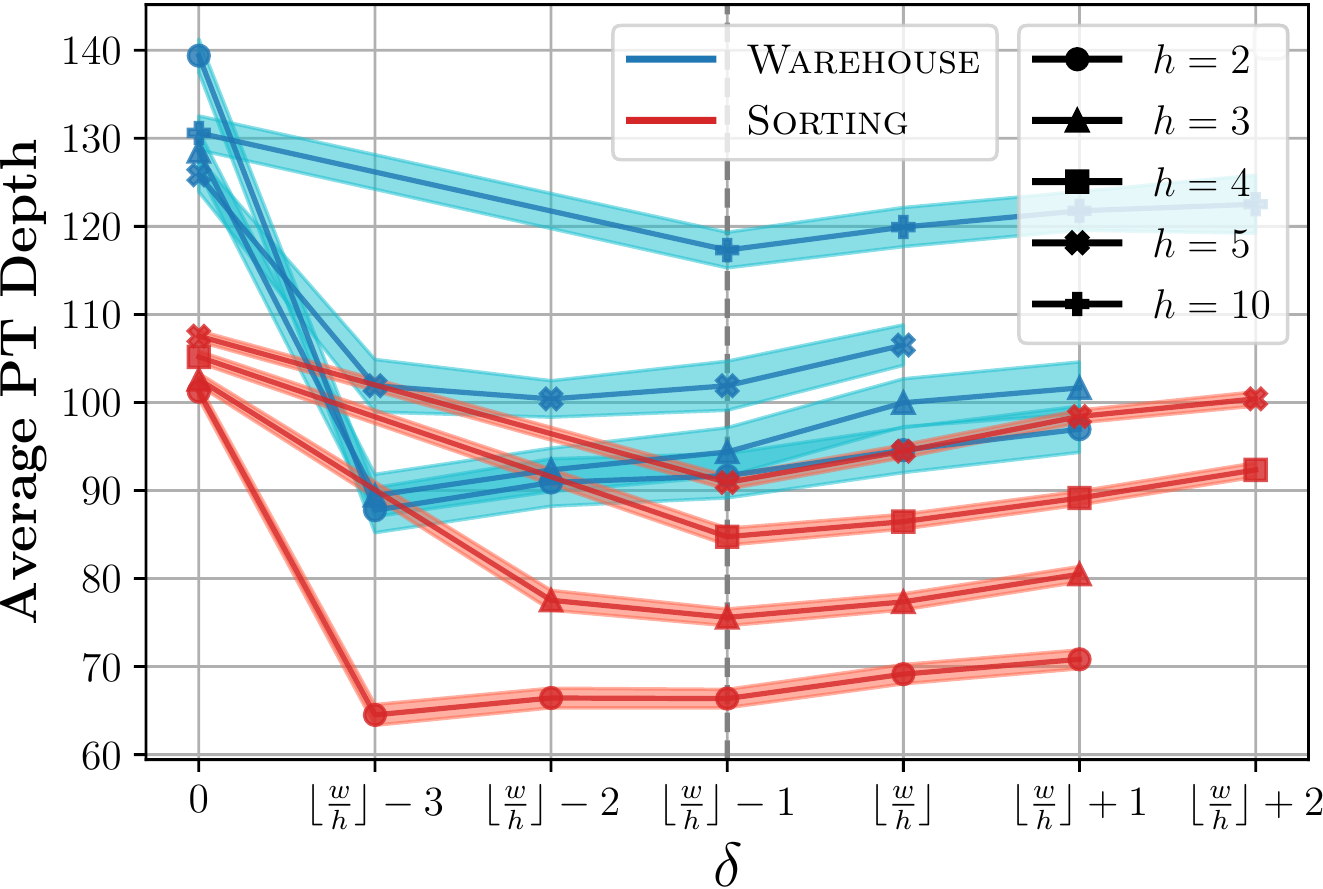} 
    \caption{Average PT depth of ${\exrhcr(\PREC_{\text{exp}})}$ for the \warehouse and \sorting domains using 150 and 300 agents, respectively for different values of $\delta$. Here, $\delta=0$ corresponds to running \rhcr without using experience. 
    }
    \label{fig:delta_experiments}
\end{figure}

We consider the effect that the lookahead parameter $\delta$ has on the performance of 
${\exrhcr(\PREC_{\text{exp}})}$.
Specifically, we fixed $w=10$ and $w=20$ for the \sorting and \warehouse environments, respectively and evaluate the average PT depth as a function of $\delta$ for different values of $h$ (\Fig \ref{fig:delta_experiments})

First, we observe that our method improves over \rhcr for every value of $\delta$ considered.
Additionally, our suggested value of $\delta=\lfloor\frac{w}{h}\rfloor-1$ yields close-to-optimal values and is a good rule of thumb in the absence of any other information.
Not surprisingly, selecting $\delta>\lfloor\frac{w}{h}\rfloor-1$ reduces performance in all of the cases tested, as it uses experience outside the planning horizon for which it was created.
Finally, selecting $\delta<\lfloor\frac{w}{h}\rfloor-1$ may be  beneficial when the replanning rate is smaller and the time window is large as the relevance of the experience seems to diminish as $\delta$ increases. For example, if $w=20, h=2$ and $\delta=\lfloor\frac{w}{h}\rfloor-1=9$, in the last query  the experience used, it' relevant for only $2$ out of the $20$ steps.

We also compare our framework to two alternatives that do not use the lookahead parameter $\delta$, as mentioned in \Sec \ref{sec:exrhcr}: (i)~running \pbs once and using the same  experience for all subsequent queries within \expbs and (ii)~using the priority set from a solution for a given query as an experience in the next query.
We used the \warehouse domain with the same parameter as in \Sec \ref{sec:exp_lmapf} and $k\in\{180,220\}$. We found that the average runtime decreases by 8\% and 21\%, respectively,  using alternative (i)~compared to \exrhcr, and by 9\% and 15\%, respectively using alternative (ii).
Additionally, the success rate dropped by 15.2\% and 11.1\% using (i), and by 10.6\% and 11.1\% using (ii).

\ignore{
We expect that increasing the scenario density, $\frac{k}{|V|}$, where $|V|$ is the number of graph vertices would also necessitate smaller values of $\delta$. That is because, as $k$ increases towards a highly-congested scenario, the number of collisions to be avoided, and consequently the depth of PT, increase, hence making it more difficult to reuse experience from one W-MAPF instance to the next.
}

\subsection{Effect of Width Limit $\ell$ on \wldfs}
\label{sec:exp_ell}

\begin{figure}[t]  
    \centering
    \includegraphics[width=0.99\linewidth]{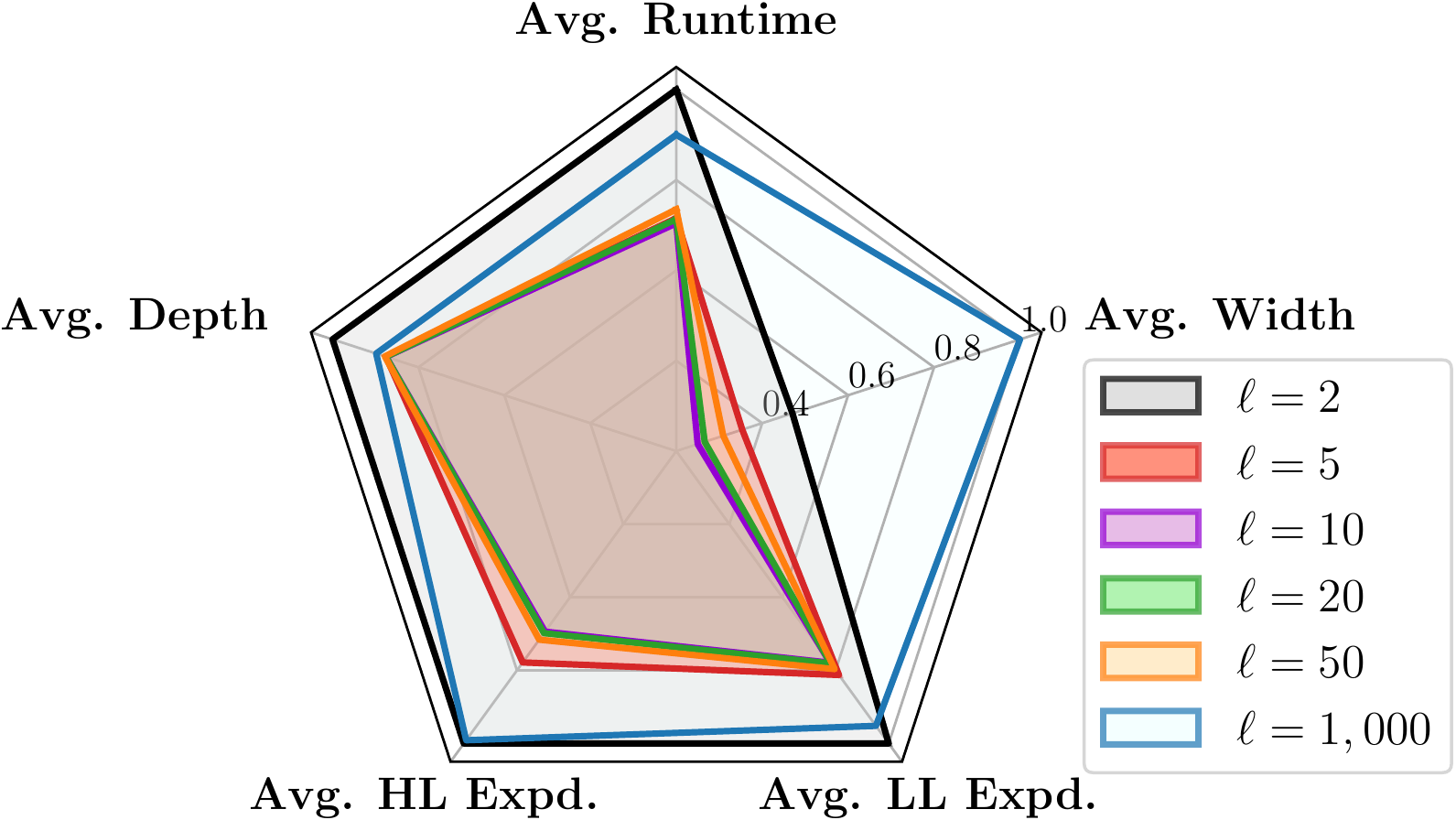} 
    \caption{Radar plot depicting the effect of~$\ell$ on various attributes of ${\exrhcr(\PREC_{\text{exp}})}$. All  values are normalized according to the maximal value attained on the axis they belong to.}
    \label{fig:wldfs_poc}
\end{figure}

We assess how the width limit parameter $\ell$ affects the performance of \expbs and consequently ${\exrhcr(\PREC_{\text{exp}})}$. 
We use
the \warehouse environment with ${w=20, h=5}$ and ${\delta=3}$ and fixed the number of agents to be $k=150$. We report average metrics (detailed shortly) for MAPF queries for different values of $\ell \in \{2, 5, 10, 20, 50, 1{,}000\}$ in \Fig \ref{fig:wldfs_poc}. 
Note that for $\ell=2$ \expbs behaves very similarly to \pbs as it
reverts to \pbs after the first backtrack. 
The setting of $\ell=1{,}000$ simulates a version wherein no fallback is taken and \wldfs is equivalent to~\dfs.

For every value of $\ell$ we logged the following attributes and report averaged values across all MAPF queries  (starting with the top attribute and going in a clockwise manner with respect to \Fig \ref{fig:wldfs_poc}): (i) average runtime; (ii) average PT width; (iii) average number of \astar node expansions in the low-level \expbs and \pbs search; (iv) average number of PT node expansions in the high-level \expbs and \pbs search; (v) average depth of PT (in case of a fallback, it consists of the sum between the depth of the \expbs and \pbs trees).
The reported values in the plot are normalized by the maximal value per attribute.

Having a small 
opportunity to reuse experience $(\ell=2)$, or allowing the search to explore the PT rooted in a seed experience indefinitely $(\ell=1{,}000)$, significantly reduces the performance of \exrhcr in comparison with the other values of~$\ell$ for all attributes indicating the benefits of \wldfs.

However, the reason why each one of these extreme values is not useful is different:
When $\ell=2$, \expbs starts to run and in many cases (roughly, $65\%$ of the times) falls back to \pbs.
This implies a small overhead caused by unnecessarily running \expbs many times.
In contrast, when $\ell=1{,}000$, \expbs rarely falls back to \pbs which may imply PT overexploration the PT when the \pbs fallback
could have found a solution quicker. 
This implies a large overhead of running \expbs incurred a small number of times. 

Finally, we consider an alternative implementation that uses a limit on the number of node expansions (rather than width). We used the \warehouse domain with the same parameter as used in \Sec \ref{sec:exp_lmapf} and $k\in\{180, 220\}$.
To estimate the number of nodes to be used as a parameter, we empirically found from previous experiments the average number of nodes expanded when exPBS succeeds in a given scenario. 
In our case, this was 99 and 209 for 180 and 220 agents, respectively. 
We evaluated a node limit of 80, 100 and 150 for 180 agents and a node limit of 150, 200 and 250 for 220 agents. 
We found that the success rate dropped by 0\%--2.6\% for 180 agents and by 5.6\%--56\% for 220 agents. 
Moreover, the runtime increased by roughly 1\%--18\% for 180 agents and by 0.7\%--10\% for 220 agents.
We note the parameter selection is more sensitive and domain-dependent compared to width limit, as mentioned in \Sec \ref{sec:expbs}.

%% file: tex/conclusions.tex
\section{Conclusions and Future Work}\label{sec:conclusion}
In this paper we described $\exrhcr$, a new approach for leveraging experience in L-MAPF instances, which allows to reduce computational effort in constituent MAPF queries by reusing priority sets from previous queries within \wmapf solvers. We demonstrated empirically that our approach can substantially improve runtime and has the potential to increase system throughput by incorporating additional agents.

Our work introduces various directions for future research. In the short run, we plan to explore approaches for systematic selection of the parameter  $\ell$, and additional heuristics for terminating and restarting the \expbs search. 

In the long run, it would be beneficial to consider advanced experience-retrieval strategies. For example, can we design an  \quot{experience database} that contains queries and their solution priorities? Here one can potentially retrieve the nearest-neighbor query under some metric to be used in \expbs. Learning-based approaches can come in handy as well by, e.g., identifying similar queries and generating experience artificially. 

Finally, the algorithms we introduced can be used to create a \emph{hierarchical approach} for solving W-MAPF problems in L-MAPF: after running \pbs we obtain a partial priority~$\PREC_{\text{exp}}$ and then compute a consistent total priority $\PREC_{\text{exp-tot}}$. We then start by running a prioritized planner using $\PREC_{\text{exp-tot}}$. In the case of failure, we fall back to \expbs  which uses~$\PREC_{\text{exp}}$ to warm-start the search and if this planner fails, we fall back to \pbs.

%% file: tex/acknowledgements.tex
\section*{Acknowledgments}
\label{sec:acknowledgements}
This research was supported in part by the Israeli Ministry of Science \& Technology grants no. 3-16079 and 3-17385, by the United States-Israel Binational Science Foundation (BSF) grants no. 2019703 and 2021643, and the Ravitz Fellowship.
The authors also thank the anonymous reviewers for their insightful comments and suggestions.

%% file: tex/appendix-new.tex
\begin{figure*}[h!]  
\centering
    \centering
    \begin{subfigure}[t]{0.32\linewidth}
        \centering
        {%
        \includegraphics[width=0.85\linewidth]{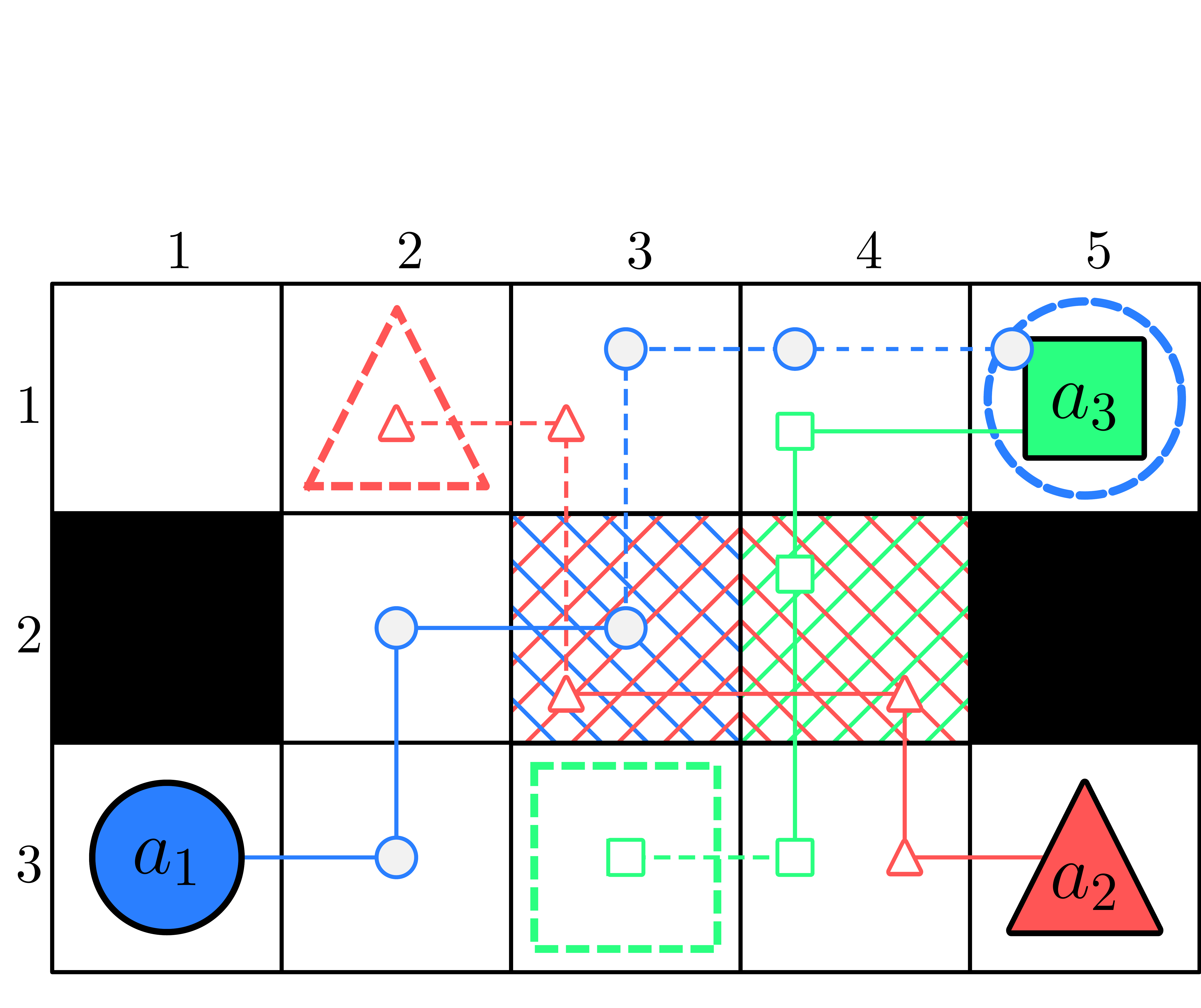} }
        \caption{Low-level search for the root node  $N_0$.} 
        \label{fig:toy_N0}
    \end{subfigure}
    \hfill
    \centering
    \begin{subfigure}[t]{0.32\linewidth}
        \centering
        {%
        \includegraphics[width=0.85\linewidth]{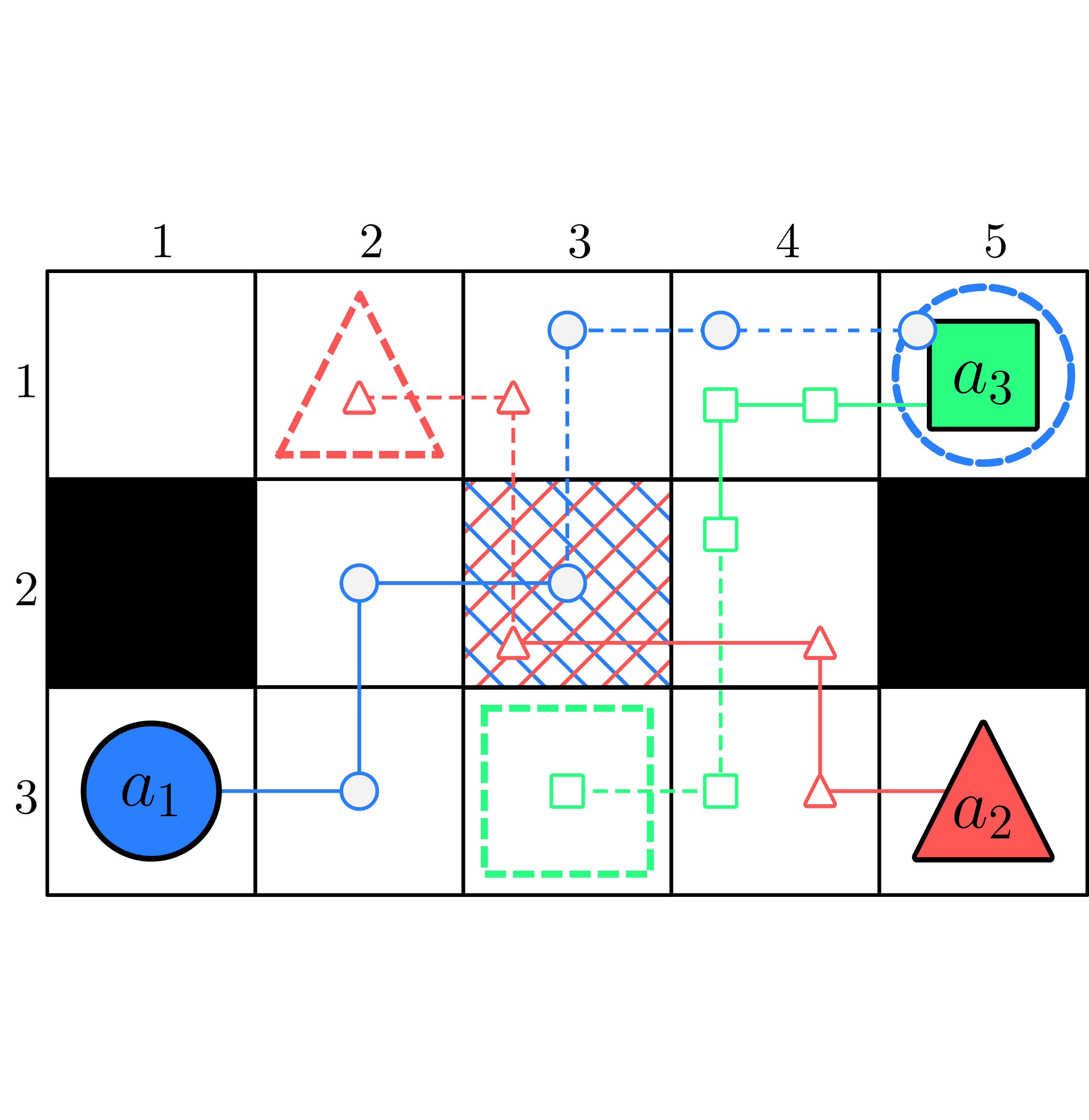} }
        \caption{Low-level search for the node $N_2$.} 
        \label{fig:toy_N2}
    \end{subfigure}
    \hfill
    \centering
    \begin{subfigure}[t]{0.32\linewidth}
        \centering
        {%
        \includegraphics[width=0.85\linewidth]{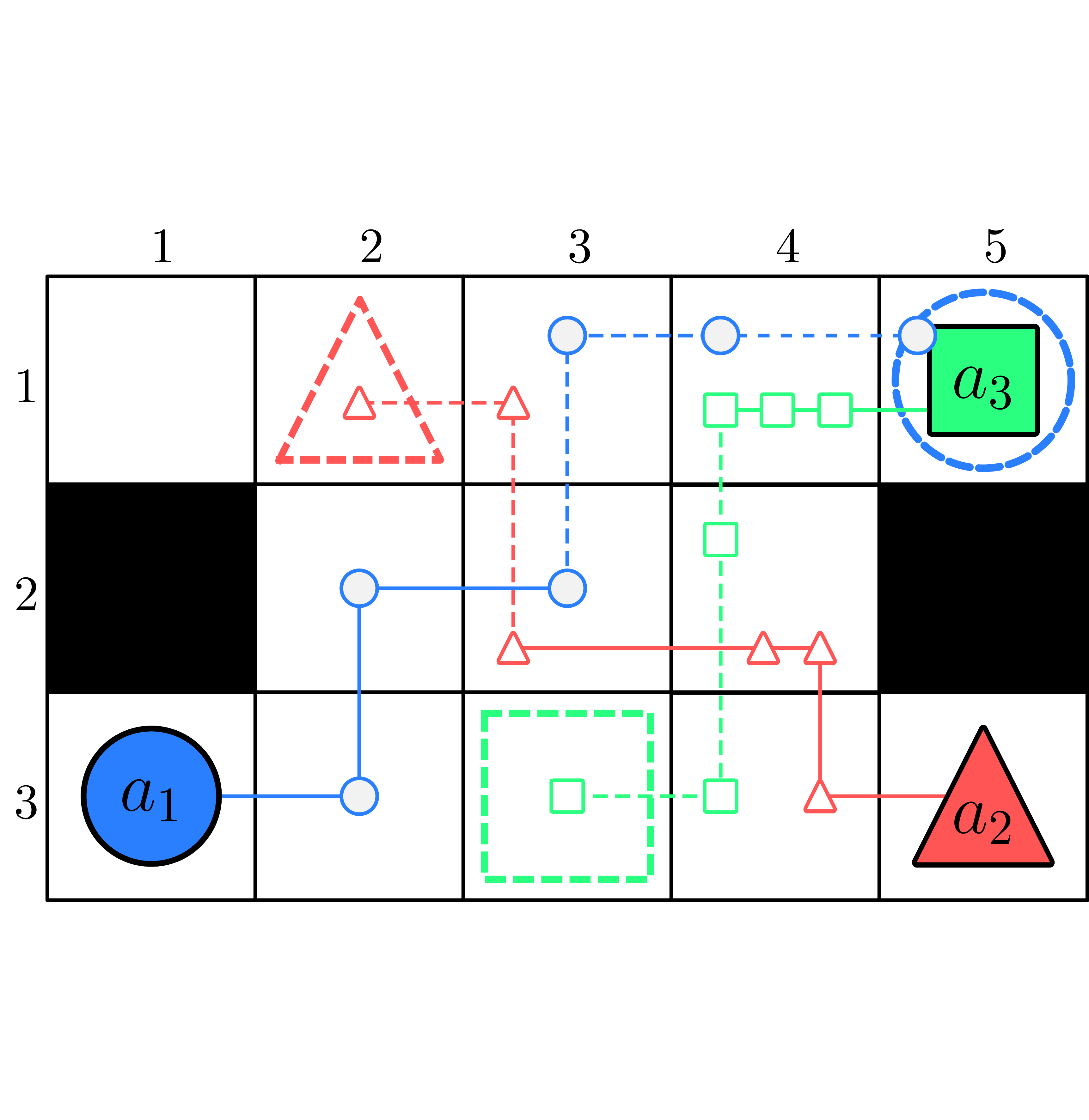} }
        \caption{Low-level search for the node $N_4$.} 
        \label{fig:toy_N4}
    \end{subfigure}
    \hfill
    \caption{Visualization of the low-level search of \pbs for W-MAPF query.}
\end{figure*}

\begin{figure}[t]  
    \centering
    \includegraphics[width=0.65\linewidth]{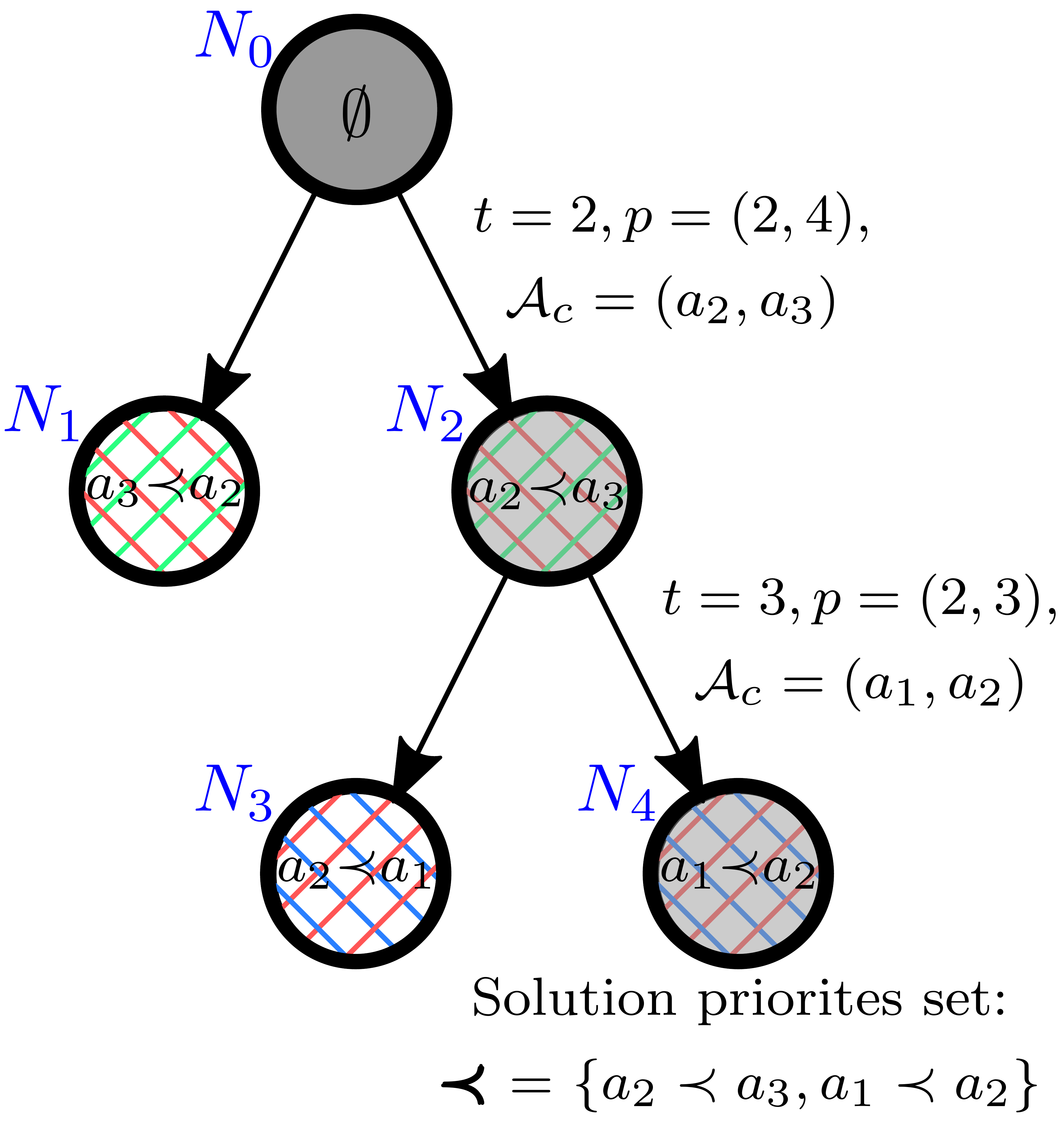} 
    \caption{\pbs search tree for W-MAPF query $q_0$.}
    \label{fig:toy_PT}
\end{figure}

\newpage

\section*{Appendix}
\appendix

We extend the explanation of the \exrhcr toy example we used in \Fig \ref{fig:toy_example}  by providing a visualization of how \pbs constructs its PT and obtains a  solution  for the first W-MAPF query $q_0$. Recall that ${w=4, h=2, \delta=2}$ and we have three agents (solid shapes) with three assigned tasks represented by goal location in dashed shapes. 

In \Fig \ref{fig:toy_PT} we visualize the final structure of the PT tree of \pbs after a solution is found. The high-level search begins in the root node~$N_0$ with the empty priority set. \Fig \ref{fig:toy_N0} illustrates the solution paths obtained by the low-level search for node~$N_0$. In particular, each agent takes its shortest path without regard to the other agents as no priorities were specified. 
The paths are:
\begin{equation*} \label{eq1}
\begin{split}
\calP   = \{ &
p_{a_1}: [(3, 1), (3, 2), (2, 2), (2, 3), (1, 3), (1, 4), (1, 5) ], \\
 & p_{a_2}: [ (3, 5), (3, 4), (2, 4), (2, 3), (1, 3), (1, 2) ], \\
 & p_{a_3}: [ (1, 5), (1, 4), (2, 4), (3, 4), (3, 3) ]  
 \}
\end{split}
\end{equation*}
where for a pair $(\cdot, \cdot)$ the values represent the row and column numbers of a cell. As shown, the first collision ${\calA_c=(a_2, a_3)}$ occurs between agent $a_2$ and agent $a_3$, at timestep $t=2$ in position $p=(2, 4)$ (this position is visualized with a criss-cross red-green pattern denoting the agents in collision).
The second collision occurs between agent $a_1$ and $a_2$ at timestep $t=3$ in position $p=(2, 4)$ (marked by blue-red criss-cross). Due to the collision, the node $N_0$ is expanded into the two nodes $N_1$ and $N_2$. As the first collision occurs between $a_2$ and $a_3$, the priority $a_3\prec a_2$ is added to the priority set of $N_1$, and $a_2\prec a_3$ to $N_2$. 

Next, we assume for the purpose of the example that \dfs chooses $N_2$ for expansion (expanded nodes are marked with  gray background) and executes the low-level search to find paths that abide to the priority set ${\PREC_{N_2}=\{a_2\prec a_3\}}$. As there are no specified priorities between $a_1$ and~$a_2$ their paths from~$N_0$ remain as-is, whereas the path of agent $a_3$ is updated and a wait action is added at position $p=(1, 4)$, as it needs to be executed while giving precedence to $a_2$ (see \Fig \ref{fig:toy_N2}). The new path of $a_3$ is $p_{a_3} = [(1, 5), (1, 4), (1, 4), (2, 4),(3, 4),(3, 3)]$.
The current state search is still a conflict between $\calA_c=(a_1, a_2)$ as detailed before.
Consequently, two child nodes $N_3$ and $N_4$ with the additional priorities $a_2\prec a_1$ and $a_1\prec a_2$, respectively, are added to the PT.

In the next step of the high-level search, \dfs picks $N_4$ for expansion, where $\PREC_{N_4} = \{a_2 \prec a_3, a_1 \prec a_2 \}$, which induces the total priority order of $a_1 \prec a_2 \prec a_3$.
Correspondingly, a low-level solution can be defined to be
\begin{equation*} \label{eq2}
\begin{split}
\calP   = \{ &
p_{a_1}: [(3, 1), (3, 2), (2, 2), (2, 3), (1, 3), (1, 4), (1, 5) ], \\
 & p_{a_2}: [ (3, 5), (3, 4),(2, 4), (2, 4), (2, 3), (1, 3), (1, 2) ], \\
 & p_{a_3}: [ (1, 5), (1, 4),(1, 4), (1, 4), (2, 4), (3, 4), (3, 3) ]
 \}.
\end{split}
\end{equation*}
As there are no conflicts for the first  $w$ timesteps, we have a found a feasible solution (\Fig \ref{fig:toy_N4}).
Note that this solution is not unique:  other priorities set can be used to solve this \wmapf, and different paths can be obtained by the low-level search depending on tie breaks.

\ignore{
\begin{figure*}[t!]  
\centering
    \begin{subfigure}[t]{1.95\columnwidth}
        \centering
        \vspace{5pt}
        \includegraphics[width=0.4\linewidth]{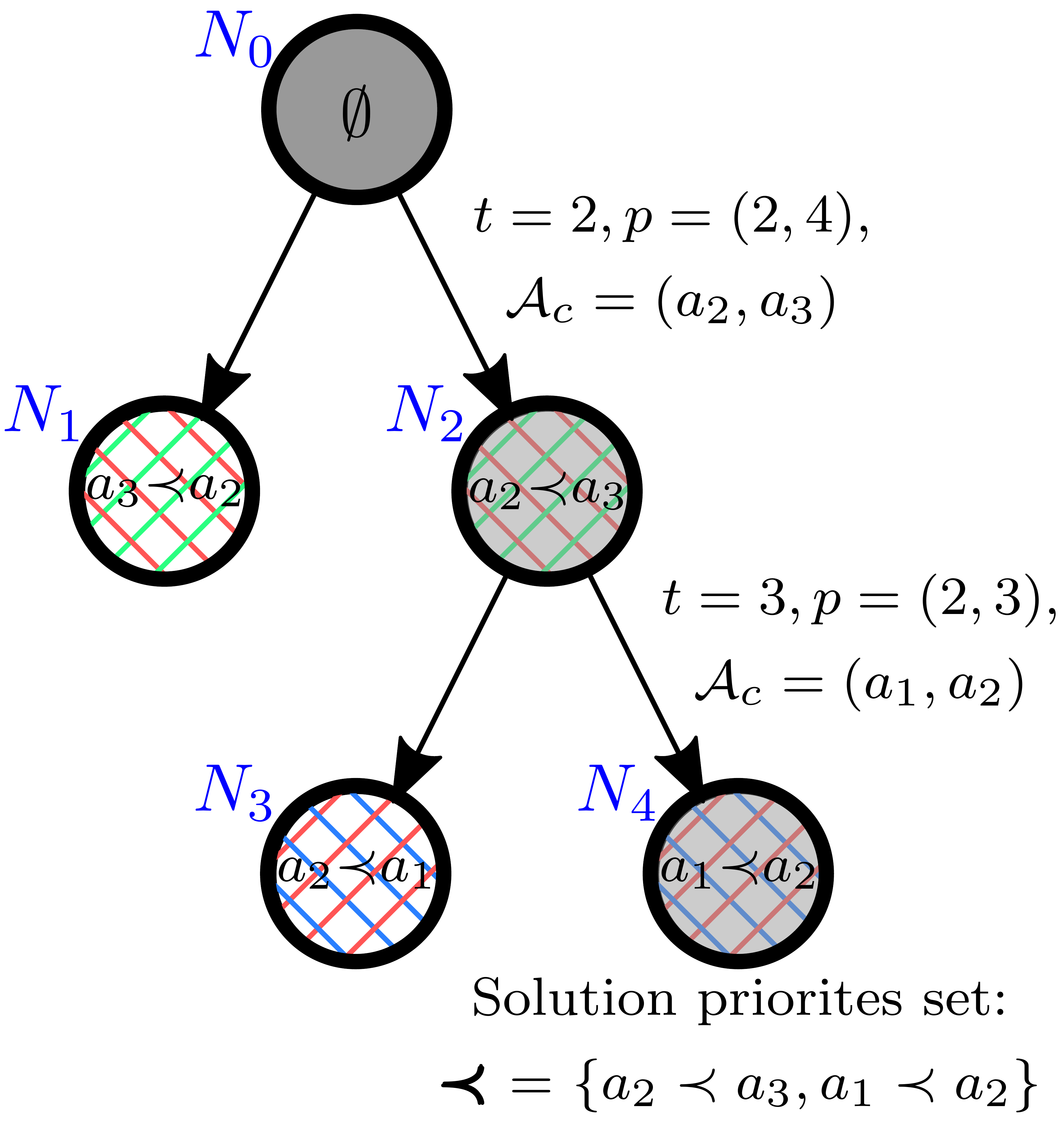} 
        \caption{\pbs search tree for W-MAPF query $q_0$.} 
        \label{fig:toy_PT}
    \end{subfigure}
    \hfill
    \centering
    \begin{subfigure}[t]{0.45\columnwidth}
        \centering
        {%
        \includegraphics[width=\linewidth]{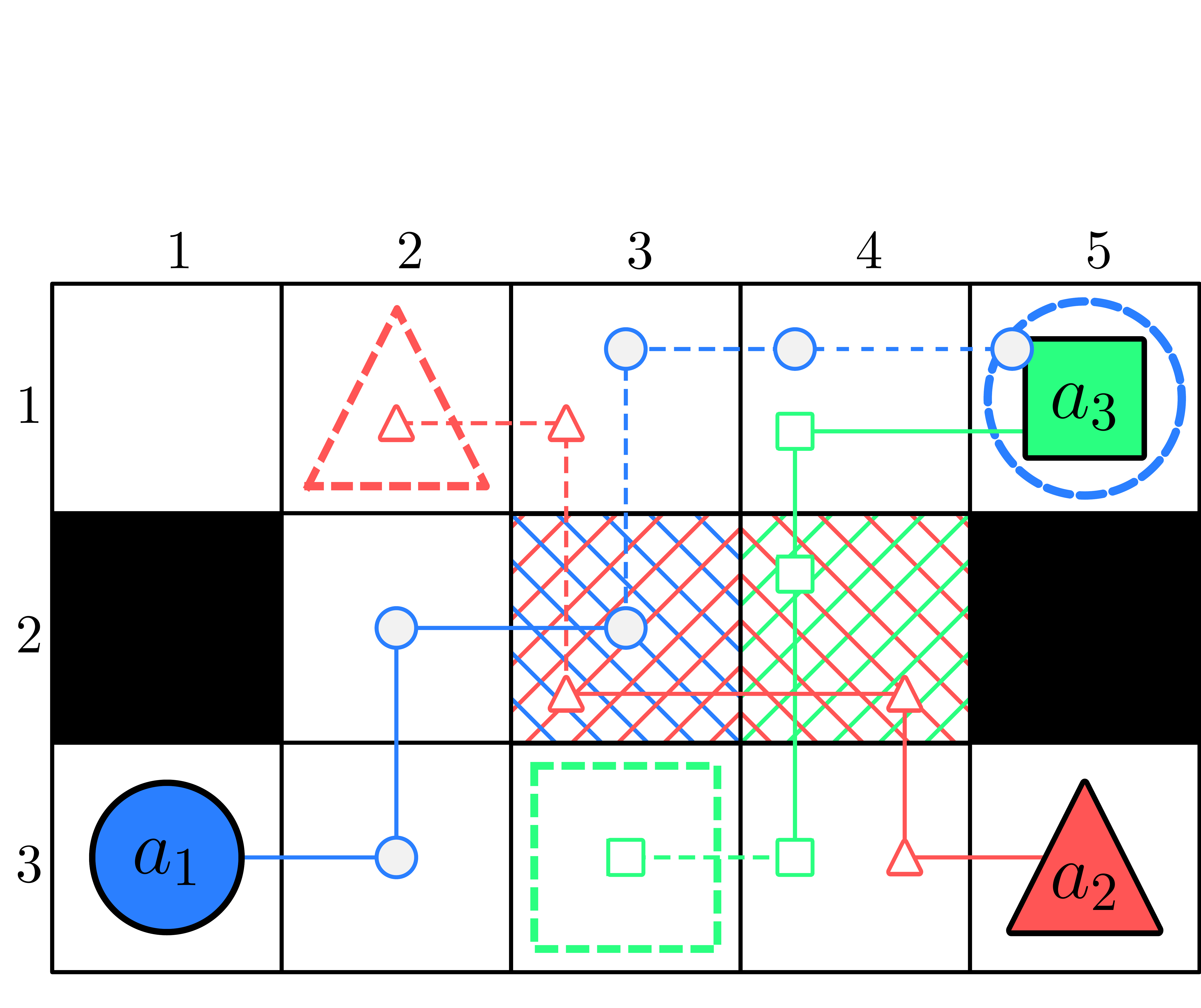} }
        \caption{Low-level search for the root node  $N_0$.} 
        \label{fig:toy_N0}
    \end{subfigure}
    \hfill
    \centering
    \begin{subfigure}[t]{0.45\columnwidth}
        \centering
        {%
        \includegraphics[width=\linewidth]{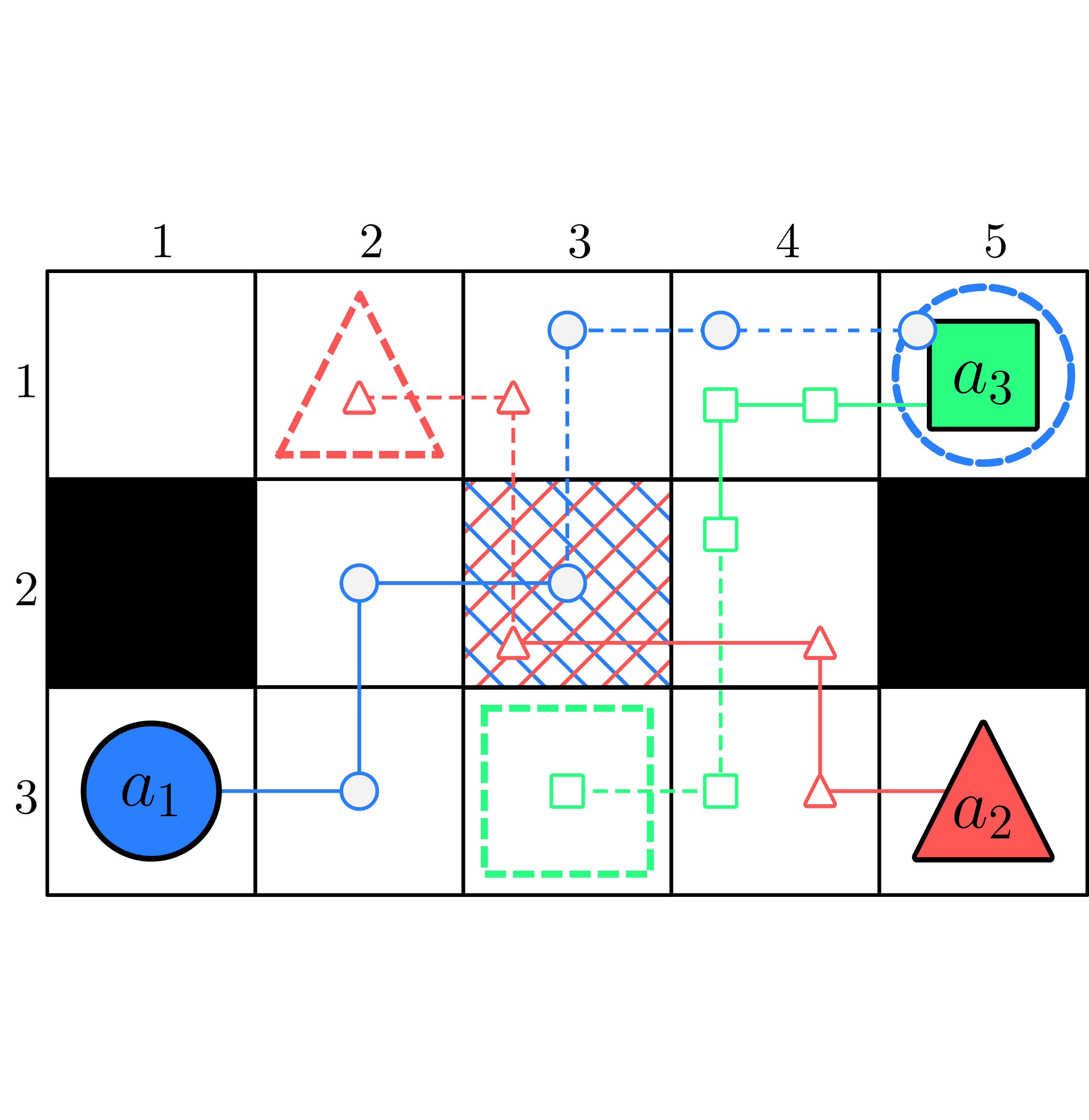} }
        \caption{Low-level search for the node $N_2$.} 
        \label{fig:toy_N2}
    \end{subfigure}
    \hfill
    \centering
    \begin{subfigure}[t]{0.45\columnwidth}
        \centering
        {%
        \includegraphics[width=\linewidth]{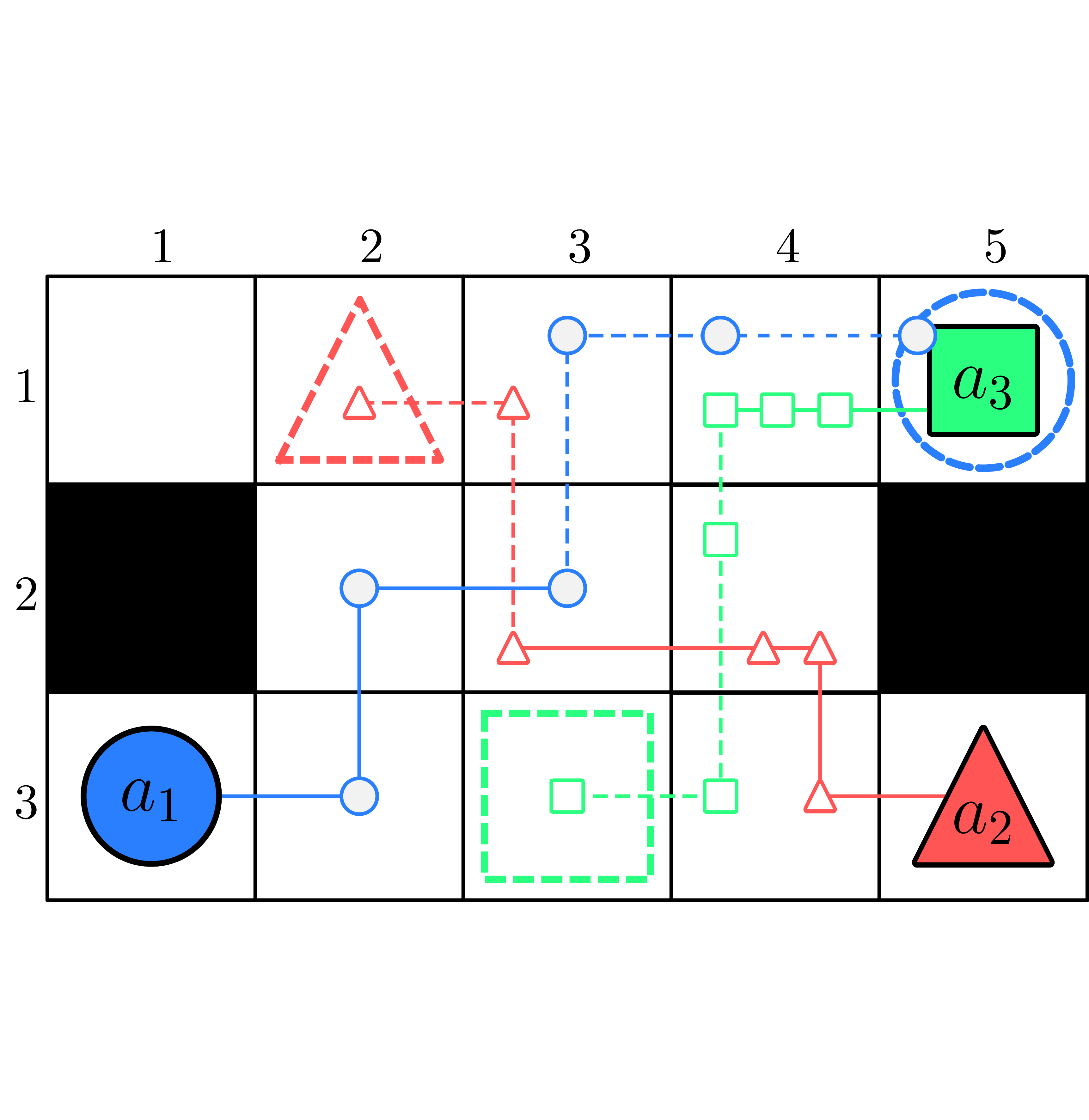} }
        \caption{Low-level search for the node $N_4$.} 
        \label{fig:toy_N4}
    \end{subfigure}
    \hfill
    \caption{Visualization of (a)~the high-level and (b,c,d)~the low-level search of \pbs on the W-MAPF query $q_0$ from the toy example.}
\end{figure*}
}

%% file: aaai22.bbl
\begin{thebibliography}{28}
\providecommand{\natexlab}[1]{#1}

\bibitem[{Barer et~al.(2014)Barer, Sharon, Stern, and
  Felner}]{barer2014suboptimal}
Barer, M.; Sharon, G.; Stern, R.; and Felner, A. 2014.
\newblock Suboptimal variants of the conflict-based search algorithm for the
  multi-agent pathfinding problem.
\newblock In \emph{{{S}o{CS}}}, 19--27.

\bibitem[{Choudhury et~al.(2021)Choudhury, Solovey, Kochenderfer, and
  Pavone}]{Choudhury.ea.21}
Choudhury, S.; Solovey, K.; Kochenderfer, M.~J.; and Pavone, M. 2021.
\newblock Efficient Large-Scale Multi-Drone Delivery using Transit Networks.
\newblock \emph{J. Artif. Intell. Res.}, 70: 757--788.

\bibitem[{Cohen and Koenig(2016)}]{cohen2016bounded}
Cohen, L.; and Koenig, S. 2016.
\newblock Bounded suboptimal multi-agent path finding using highways.
\newblock In \emph{{{IJCAI}}}, 3978--3979.

\bibitem[{Coleman et~al.(2015)Coleman, {\c{S}}ucan, Moll, Okada, and
  Correll}]{coleman2015experience}
Coleman, D.; {\c{S}}ucan, I.~A.; Moll, M.; Okada, K.; and Correll, N. 2015.
\newblock Experience-based planning with sparse roadmap spanners.
\newblock In \emph{{{ICRA}}}, 900--905.

\bibitem[{Dayan et~al.(2021)Dayan, Solovey, Pavone, and Halperin}]{Dayan.ea.21}
Dayan, D.; Solovey, K.; Pavone, M.; and Halperin, D. 2021.
\newblock Near-Optimal Multi-Robot Motion Planning with Finite Sampling.
\newblock In \emph{{{ICRA}}}, 9190--9196.

\bibitem[{Demaine et~al.(2019)Demaine, Fekete, Keldenich, Meijer, and
  Scheffer}]{Demaine.ea.19}
Demaine, E.~D.; Fekete, S.~P.; Keldenich, P.; Meijer, H.; and Scheffer, C.
  2019.
\newblock Coordinated Motion Planning: Reconfiguring a Swarm of Labeled Robots
  with Bounded Stretch.
\newblock \emph{{SIAM} J. on Comput.}, 48(6): 1727--1762.

\bibitem[{Gordon, Filmus, and Salzman(2021)}]{Gordon.ea.21}
Gordon, O.; Filmus, Y.; and Salzman, O. 2021.
\newblock Revisiting the Complexity Analysis of Conflict-Based Search: New
  Computational Techniques and Improved Bounds.
\newblock In \emph{{{S}o{CS}}}, 64--72.

\bibitem[{H{\"o}nig et~al.(2019)H{\"o}nig, Kiesel, Tinka, Durham, and
  Ayanian}]{honig2019persistent}
H{\"o}nig, W.; Kiesel, S.; Tinka, A.; Durham, J.~W.; and Ayanian, N. 2019.
\newblock Persistent and robust execution of {MAPF} schedules in warehouses.
\newblock \emph{{{RA-L}}}, 4(2): 1125--1131.

\bibitem[{Huang, Koenig, and Dilkina(2021)}]{Huang.ea.21}
Huang, T.; Koenig, S.; and Dilkina, B. 2021.
\newblock Learning to Resolve Conflicts for Multi-Agent Path Finding with
  Conflict-Based Search.
\newblock In \emph{{AAAI}}, 11246--11253. {AAAI} Press.

\bibitem[{Kaduri, Boyarski, and Stern(2020)}]{Kaduri.ea.20}
Kaduri, O.; Boyarski, E.; and Stern, R. 2020.
\newblock Algorithm Selection for Optimal Multi-Agent Pathfinding.
\newblock In \emph{{{ICAPS}}}, 161--165.

\bibitem[{Li et~al.(2020)Li, Gange, Harabor, Stuckey, Ma, and
  Koenig}]{li2020new}
Li, J.; Gange, G.; Harabor, D.; Stuckey, P.~J.; Ma, H.; and Koenig, S. 2020.
\newblock New techniques for pairwise symmetry breaking in multi-agent path
  finding.
\newblock In \emph{{{ICAPS}}}, volume~30, 193--201.

\bibitem[{Li et~al.(2019{\natexlab{a}})Li, Harabor, Stuckey, Felner, Ma, and
  Koenig}]{li2019disjoint}
Li, J.; Harabor, D.; Stuckey, P.~J.; Felner, A.; Ma, H.; and Koenig, S.
  2019{\natexlab{a}}.
\newblock Disjoint splitting for multi-agent path finding with conflict-based
  search.
\newblock In \emph{{{ICAPS}}}, volume~29, 279--283.

\bibitem[{Li et~al.(2019{\natexlab{b}})Li, Harabor, Stuckey, Ma, and
  Koenig}]{li2019symmetry}
Li, J.; Harabor, D.; Stuckey, P.~J.; Ma, H.; and Koenig, S. 2019{\natexlab{b}}.
\newblock Symmetry-breaking constraints for grid-based multi-agent path
  finding.
\newblock In \emph{{AAAI}}, volume~33, 6087--6095.

\bibitem[{Li, Ruml, and Koenig(2021)}]{li2021eecbs}
Li, J.; Ruml, W.; and Koenig, S. 2021.
\newblock EECBS: A bounded-suboptimal search for multi-agent path finding.
\newblock In \emph{{AAAI}}.

\bibitem[{Li et~al.(2021)Li, Tinka, Kiesel, Durham, Kumar, and
  Koenig}]{li2021lifelong}
Li, J.; Tinka, A.; Kiesel, S.; Durham, J.~W.; Kumar, T. K.~S.; and Koenig, S.
  2021.
\newblock Lifelong Multi-Agent Path Finding in Large-Scale Warehouses.
\newblock In \emph{{AAAI}}, 11272--11281.

\bibitem[{Liu et~al.(2019)Liu, Ma, Li, and Koenig}]{liu2019task}
Liu, M.; Ma, H.; Li, J.; and Koenig, S. 2019.
\newblock Task and path planning for multi-agent pickup and delivery.
\newblock In \emph{{{AAMAS}}}, 1152--1160.

\bibitem[{Ma et~al.(2019{\natexlab{a}})Ma, Harabor, Stuckey, Li, and
  Koenig}]{ma2019searching}
Ma, H.; Harabor, D.; Stuckey, P.~J.; Li, J.; and Koenig, S. 2019{\natexlab{a}}.
\newblock Searching with consistent prioritization for multi-agent path
  finding.
\newblock In \emph{{AAAI}}, volume~33, 7643--7650.

\bibitem[{Ma et~al.(2019{\natexlab{b}})Ma, H{\"{o}}nig, Kumar, Ayanian, and
  Koenig}]{Ma.Honig.ea.19}
Ma, H.; H{\"{o}}nig, W.; Kumar, T. K.~S.; Ayanian, N.; and Koenig, S.
  2019{\natexlab{b}}.
\newblock Lifelong Path Planning with Kinematic Constraints for Multi-Agent
  Pickup and Delivery.
\newblock In \emph{{AAAI}}, 7651--7658.

\bibitem[{Ma et~al.(2017)Ma, Li, Kumar, and Koenig}]{ma2017lifelong}
Ma, H.; Li, J.; Kumar, T. K.~S.; and Koenig, S. 2017.
\newblock Lifelong multi-agent path finding for online pickup and delivery
  tasks.
\newblock In \emph{{{AAMAS}}}, 837--845.

\bibitem[{Ma et~al.(2016)Ma, Tovey, Sharon, Kumar, and Koenig}]{Ha.ea.16}
Ma, H.; Tovey, C.~A.; Sharon, G.; Kumar, T. K.~S.; and Koenig, S. 2016.
\newblock Multi-Agent Path Finding with Payload Transfers and the
  Package-Exchange Robot-Routing Problem.
\newblock In \emph{{AAAI}}, 3166--3173.

\bibitem[{Salzman and Stern(2020)}]{salzman2020research}
Salzman, O.; and Stern, R. 2020.
\newblock Research challenges and opportunities in multi-agent path finding and
  multi-agent pickup and delivery problems.
\newblock In \emph{AAMAS}, 1711--1715.

\bibitem[{Sharon et~al.(2015)Sharon, Stern, Felner, and
  Sturtevant}]{sharon2015conflict}
Sharon, G.; Stern, R.; Felner, A.; and Sturtevant, N.~R. 2015.
\newblock Conflict-based search for optimal multi-agent pathfinding.
\newblock \emph{Artificial Intelligence}, 219: 40--66.

\bibitem[{Sharon et~al.(2013)Sharon, Stern, Goldenberg, and
  Felner}]{sharon2013increasing}
Sharon, G.; Stern, R.; Goldenberg, M.; and Felner, A. 2013.
\newblock The increasing cost tree search for optimal multi-agent pathfinding.
\newblock \emph{Artificial Intelligence}, 195: 470--495.

\bibitem[{Silver(2005)}]{silver2005cooperative}
Silver, D. 2005.
\newblock Cooperative pathfinding.
\newblock \emph{Artificial Intelligence and Interactive Digital Entertainment},
  1: 117--122.

\bibitem[{Stern et~al.(2019)Stern, Sturtevant, Felner, Koenig, Ma, Walker, Li,
  Atzmon, Cohen, Kumar et~al.}]{stern2019benchmarks}
Stern, R.; Sturtevant, N.~R.; Felner, A.; Koenig, S.; Ma, H.; Walker, T.~T.;
  Li, J.; Atzmon, D.; Cohen, L.; Kumar, T.~S.; et~al. 2019.
\newblock Multi-agent pathfinding: Definitions, variants, and benchmarks.
\newblock In \emph{{{S}o{CS}}}, 151--158.

\bibitem[{Surynek et~al.(2016)Surynek, Felner, Stern, and
  Boyarski}]{Surynek.ea.16}
Surynek, P.; Felner, A.; Stern, R.; and Boyarski, E. 2016.
\newblock Efficient {SAT} approach to multi-agent path finding under the sum of
  costs objective.
\newblock In \emph{ECAI}, 810--818. IOS Press.

\bibitem[{Wagner and Choset(2015)}]{wagner2015subdimensional}
Wagner, G.; and Choset, H. 2015.
\newblock Subdimensional expansion for multirobot path planning.
\newblock \emph{Artificial Intelligence}, 219: 1--24.

\bibitem[{Yu(2016)}]{Yu.16}
Yu, J. 2016.
\newblock Intractability of Optimal Multirobot Path Planning on Planar Graphs.
\newblock \emph{{{RA-L}}}, 1(1): 33--40.

\end{thebibliography}
